\documentclass[mathleft]{an}
\usepackage{graphicx}
\usepackage{times}
\begin{document}

\Pagespan{000}{000}
\Yearpublication{0000}%
\Yearsubmission{2008}%
\Month{00}%
\Volume{000}%
\Issue{00}%

\title{A revisit to agglomerates of early-type {\em Hipparcos} stars}

\author{J. A. Caballero\inst{1}\fnmsep\thanks{Corresponding author:
  Jos\'e A. Caballero, investigador Juan de la Cierva at the UCM
  (\email{\tt caballero@astrax.fis.ucm.es}).\newline}
  \and L. Dinis\inst{2}}
\titlerunning{A revisit to agglomerates of early-type {\em Hipparcos} stars}
\authorrunning{J. A. Caballero \& L. Dinis}
\institute{
Dpto. de Astrof\'{\i}sica y Ciencias de la Atm\'osfera, Facultad de
Ciencias F\'{\i}sicas, Universidad Complutense de Madrid, E-28040 Madrid, Spain
\and 
Grupo Interdisciplinar de Sistemas Complejos (GISC) and Dpto. de F\'{\i}sica
At\'omica, Molecular y Nuclear, Universidad Complutense de Madrid, E-28040
Madrid, Spain}

\received{26 Jun 2008}
\accepted{30 Jun 2008}
\publonline{.. Aug 2008}

\keywords{stars: early-type -- Galaxy: open clusters and associations: general
-- open clusters and associations: individual (Ori~OB1b, P~Puppis) --
methods: data analysis -- astronomical data bases: miscellaneous} 

\abstract{%
We study the spatial structure and sub-structure of regions rich in {\em
Hipparcos} stars with blue $B_T-V_T$ colours.
These regions, which comprise large stellar complexes, OB associations, and
young open clusters, are tracers of on-going star formation in the Galaxy.
The DBSCAN (Density-Based Spatial Clustering of Applications with Noise) data
clustering algorithm is used to look for spatial overdensities of early-type 
stars.  
Once an overdensity, ``agglomerate'', is identified, we carry out a data and
bibliographic compilation of their star member candidates.
The actual membership in agglomerate of each early-type star is studied based
on its heliocentric distance, proper motion, and previous spectro-photometric
information.
We identify 35 agglomerates of early-type {\em Hipparcos} stars.
Most of them are associated to previously known clusters and OB associations.
The previously unknown P~Puppis agglomerate is subject of a dedicated study with
Virtual Observatory tools. 
It is actually a new, nearby, young open cluster ($d \sim$ 470\,pc, age $\sim$
20\,Ma) with a clear radial density gradient. 
We list P~Puppis and other six agglomerates (including NGC~2451~A, vdBH~23,
and Trumpler~10) as new sites for substellar searches because of their youth,
closeness, and spatial density.
We investigate in detail the sub-structure in the Orion, CMa-Pup and Pup-Vel
OB~complexes (``super-agglomerates'').  
We confirm or discover some stellar overdensities in the Orion complex, like
the 25~Ori group, the Horsehead region (including the $\sigma$~Orionis cluster),
and the $\eta$~Orionis agglomerate.    
Finally, we derive accurate parallactic distances to the Pleiades, NGC~2451~A,
and IC~2391, describe several field early-type stars at $d <$ 200\,pc, and
discuss the incompleteness of our search.}

\maketitle

\section{Introduction}
\label{introduction}

The works by Pannekoek (1929), Ambartsumian (1947) and Ruprecht (1966) were,
according to de~Zeeuw et~al. (1999), among the most important ones for the
understanding of the OB associations in the early pre-{\em Hipparcos} era.
We also quote the comprehensive review by Blaauw (1964).
OB associations are very young regions in the Galaxy containing high-mass
O- and B-type stars.
The ultraviolet radiation injected into the intra-association medium by these
luminous blue stars play an important r\^ole on the subsequent low-mass star
formation process.
The study of OB associations is, therefore, necessary for answering critical
questions in Astrophysics, like `how does the fragmentation of a primordial
molecular cloud take place', `which is the shape of the initial mass function',
`how is the dynamical evolution at the first stages of a recently-born open
cluster', `where are the spiral arms of our Galaxy', or `do early-type stars
inhibit or facilitate the formation of substellar objects'.   

After the {\em magnus opus} by de~Zeeuw et~al. (1999), who carried out a
comprehensive census of the stellar content of nearby ($d <$ 1\,kpc) OB
associations, based on {\em Hipparcos} positions, proper motions, and
parallaxes, and prior to launch of the European Space Agency (ESA) mission {\em
GAIA}, very few ``fresh'' discoveries can be achieved on OB associations.
The work by de~Zeeuw et~al. (1999) is not, however, the only search for star
clusters and OB associations from the {\em Hipparcos} data: Platais,
Kozhurina-Platais \& \linebreak
van~Leeuwen (1998) and Robichon et~al. (1999) had previously done it (see also a
posterior work by Baumgardt, Dettbarn \& Wielen~[2000]).  

The superior capabilities and results of the ESA astrometry mission {\em
Hipparcos} (Perryman et~al. 1997) have allowed astronomers to go on the
characterization of the Galactic stellar populations in general, and of OB
associations in particular.
There have been other works with the {\em Hipparcos} catalogue aimed at studying
particular regions containing early-type stars (Baumgardt 1998; de Bruijne 1999;
Subramaniam \& Bhatt 2000) or the Galactic structure attending to the spatial
distribution of such stars (Comer\'on, Torra \& G\'omez 1998;
Ma\'{\i}z-Apell\'aniz 2001; Schr\"oder et~al. 2004).
We all place our trust in the ``fresh'' discoveries on OB associations that will
arise in the near future with the overwhelming {\em GAIA} dataset.
It is expected, however, that the {\em GAIA} mission will be launched in the
second half of~2011.

In the interim, we can still explore the OB associations with the {\em
Hipparcos} catalogue and different aims, tools, and reductions of the raw data.
On the one hand, in the current work, we have used the late re-reduction of the
{\em Hipparcos} data by van~Leeuwen (2007a), who has obtained improvements by up
to a factor 4 in the astrometric accuracies for particular bright stars (see
also details on the validation of the new {\em Hipparcos} reduction in
van~Leeuwen [2007b]).
On the other hand, we have applied a novel (in Galactic astronomy) clustering
analysis to part a dataset into subsets (``clusters'' or ``agglomerates'').
Through {\em data clustering}, objects with a common distinguishing feature are
classified into different groups. 
In our case, the common quality is just the spatial location of {\em Hipparcos}
stars with blue colours (i.e. with early spectral types).
That is, our aim is to study the structure (and super- and sub-structure) of
Galactic OB associations, and derive some basic properties.

\section{Analysis}
\label{analysis}

\begin{figure*}
\centering
\includegraphics[width=0.48\textwidth]{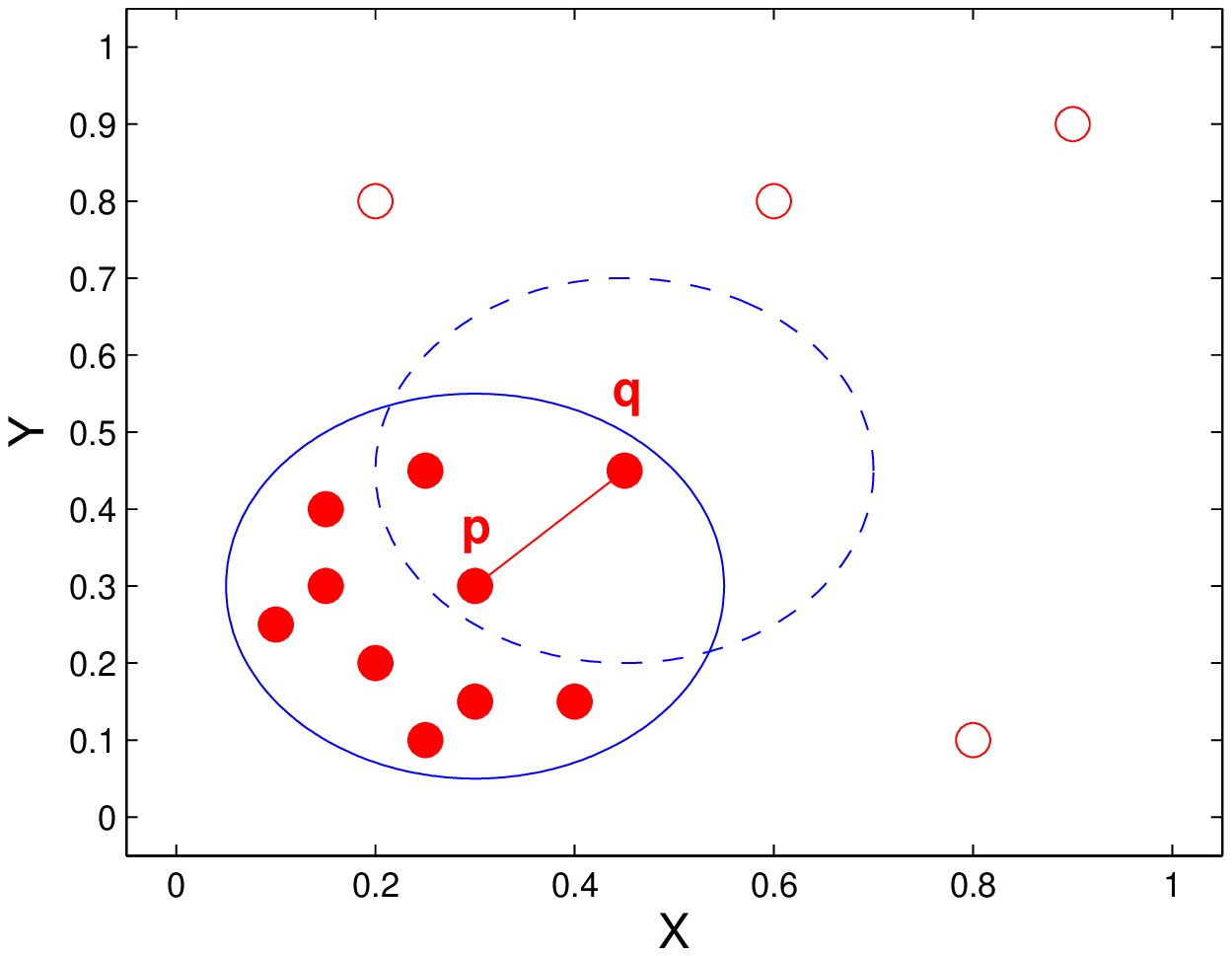} 
\includegraphics[width=0.48\textwidth]{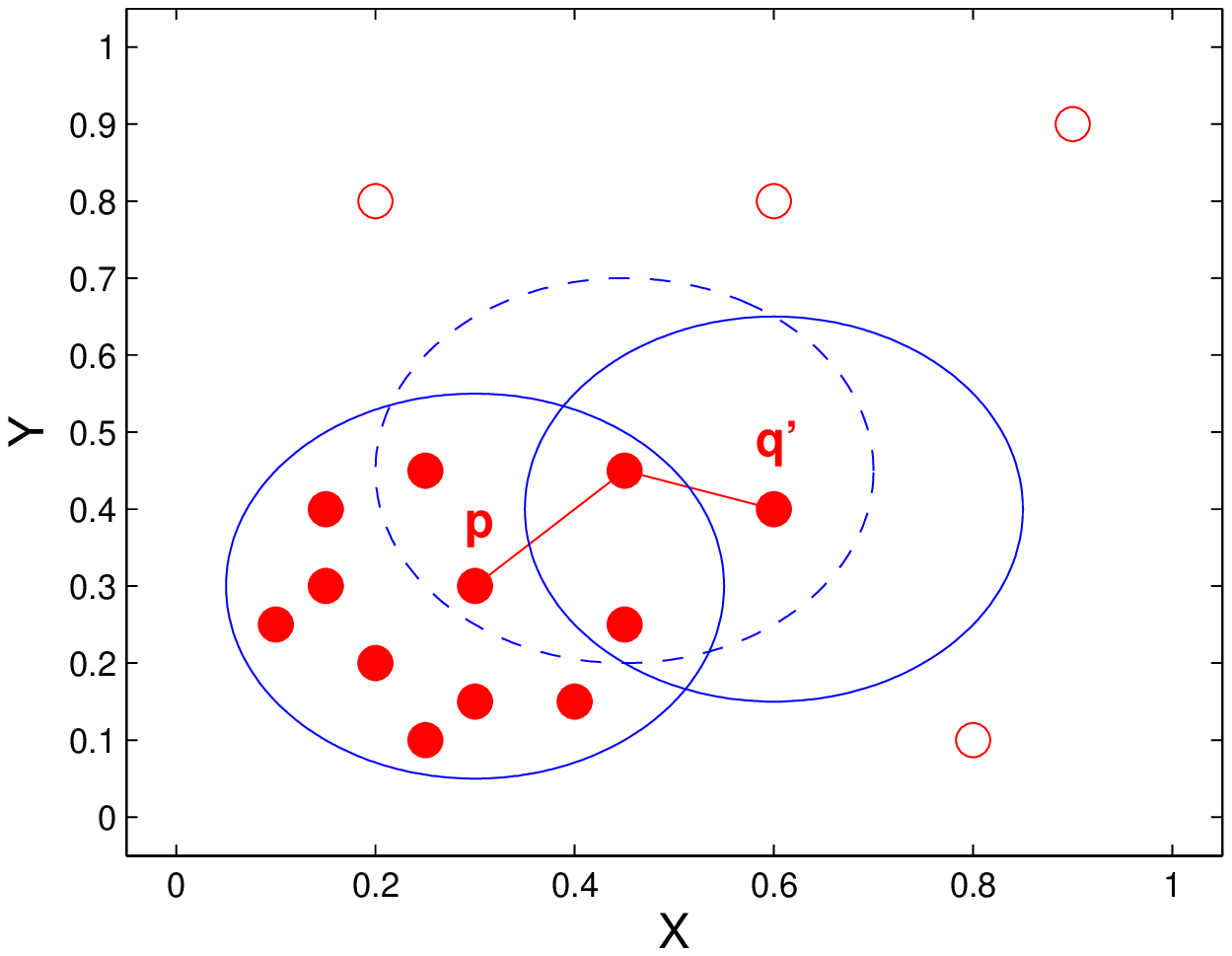} 
\caption{A core point $p$ and a border point $q$ ({\em left} window) and a point
$q'$ density-reachable from $p$ through an intermediate (core) point ({\em
right} window) for $N_{\rm MinPts}$ = 4 and $R_\epsilon$ = 0.25 in arbitrary
units. 
Filled symbols represent cluster points.
Colour versions of all our figures are available in the electronic publication.}   
\label{theluis}
\end{figure*}

\subsection{Data clustering}

\subsubsection{Types of clustering}
\label{typesofclustering}

In short, data clustering is a common technique for statistical data analysis
with the aim to classificate elements into different groups or ``clusters''.
It is widely used in some scientific disciplines, like Computational Biology
(for gene finding or plant systematics), market research (for partitioning a
population of consumers into market segments), or pattern recognition in machine
learning (according to the clustering illusion, the natural human ``see patterns
where actually none exists'').
There are several types of data clustering algorithms (and classifications, as
well).
The simplest categorization splits the clustering algorithms into hierarchical
and partitional ones;
hierarchical algorithms can be, in their turn, divisive or agglomerative.
The similarity of two elements within a cluster is quantified with a distance
measure, that is, in most of the cases, the Euclidean distance (there are also
other distances, like the Manhattan, Mahalanobis, and Hamming ones).
An agglomerative hierarchical clustering algorithm begins with each element as a
separate cluster and, through comparison with other elements, progessively
merges them into clusters. 
The single linkage clustering (``nearest neighbour'') is a simple
agglomerative~method.

\subsubsection{Clustering in Astrophysics}

Numerous astronomers and cosmologists have investigated hierarchical {\em
extragalactic} clusterings (see a good example in White \& Frenk [1991]). 
Different algorithms to find \linebreak 
groups, clusters, and superclusters of galaxies were early presented by, e.g.,
Turner \& Gott (1976), Huchra \& Geller (1982), Press \& Davis (1982), Einasto
et~al. (1984), Barrow, Bhavsar \& Sonoda (1985), Maddox et~al. (1990), and
Mart\'{\i}nez et~al. (1990). 
The used algorithms consisted in two-point angular correlation functions,
minimal spanning trees, multifractal measures, percolations, an the well-tested
friends-of-friends (FoF) algorithm.
More recent clustering analyses, particularly using data from the Sloan Digital
Sky Survey and FoF-like algorithms, have been carried out by
Merch\'an \& Zandivarez (2002), Berlind et~al. (2006), Weinmann et~al. (2006),
or Tago et~al. (2008). 
Besides, Uchihori et~al. (2000) looked for clusters of extremely high energy
cosmic rays in the northern sky, which are in generally ascribed to
extragalactic clusters.

At shorter distances, but still out of our Milky Way, there have been searches
of OB associations using data
clustering, mostly the Path Linkage Criterion, in the Magellanic System
(Battinelli 1991; Bica \& Schmitt 1995; Gouliermis et~al. 2000), the Andromeda
galaxy (Magnier et~al. 1993; Battinelli, Efremov \& Magnier 1996), and in other
galaxies in the Local Group (Demers et~al. 1995; Ivanov 1996; Pietrzy\'nski
et~al. 2001).  
Although it is obvious to extrapolate the use of such automatic procedures to
search for clusters in our own Galaxy, there have been very few works on this
topic.
First, Mel'nik \& Efremov (1995) used the list of OB stars of Blaha \&
Humphreys (1989) as their input catalogue and the Path Linkage Criterion (that
is a hierarchical agglomerative algorithm) to look for Galactic OB associations.
Secondly, Reyl\'e \& Robin (2002) used the same method to search for (embedded)
star clusters close to the Galacic plane with the Point Source Catalogue of the
DENIS~survey.

The spaghetti method used by de~Zeeuw et~al. (1999) and Hoogerwerf \& Aguilar
(1999) to identify OB associations and nearby moving groups, respectively, the
3-D and 2-D wavelet analysis carried out by Chereul, Cr\'ez\'e \& Bienaym\'e
(1998, 1999) and Skuljan, Heranshaw \& Cottrell (1999), and other maximum
likelihood approaches (Chen et~al. 1997; Asiain et~al. 1999) can also be
considered as uncommon examples of data clustering in the Galaxy.
However, all those methods require, in general, a complex mathematical apparatus
and information on the star proper motion and, sometimes, age.
At the long heliocentric distances where most of the OB associations are found
($d \ge$ 200\,pc), their stellar populations are difficult no disentangle
from the background solely based on proper motions, that are typically very low
($\mu \le$ 10\,mas\,a$^{-1}$).
The data clustering algorithm that we use in this paper resembles the
FoF {\em extragalactic} algorithm, that is much easier to
implement, and only requires as input the 2-D spatial coordinates ($\alpha$,
$\delta$) of the stars.
Uncertainties in the parallax determination of the farthest {\em Hipparcos}
stars may prevent the use of a 3-D FoF-like algorithm (this task will be
performed in the future with {\em GAIA}~data).

\subsubsection{Our clustering algorithm}
\label{theclusteringalgorithm}

In this section, we will use the term ``cluster'' to define the groupings of
stars obtained by the algorithm, to use the standard nomenclature in data
clustering.  
However, in the following sections, we will use the term ``agglomerate'' for
the different groupings of stars that the algorithm returns. 
This choice is to avoid confusion with the term ``(star) cluster''. 
Other expressions that we must avoid using are ``(moving) group'' and ``(OB)
association''.
Obviously, we expect the clustering algorithm to provide agglomerates of
early-type stars that ultimately belong to a star cluster, a moving group, or an
OB association, but their membership can only be confirmed after a follow-up of
the individual stars.  

Among the available algorithms for data clustering, we chose DBSCAN
(Density-Based Spatial Clustering of Applications with Noise) for its
ability to recognise clusters of arbitrary shape in a database with ``noise'',
that is, a number of points that do not necessarily belong to any cluster. 
Recalling Section~\ref{typesofclustering}, DBSCAN is an Euclidean agglomerative
hierarchical clustering algorithm. 
We will just outline here the fundamental aspects of density-based clustering
and describe briefly the algorithm. 
A more detailed description of DBSCAN, including a pseudo-code, can be found in
Ester et~al~(1996). 

The notion of cluster in DBSCAN resides in the fact that the density inside a
cluster is considerably higher than outside it, in the noise. 
This situation clearly resembles our problem, that consists in picking out
densely packed groups of stars in an apparently random 2-D spatial distribution.
To ensure that the clusters recognised by the algorithm are sufficiently dense,
one may na\"{\i}vely require that there are at least a minimum number of points
($N_{\rm MinPts}$) in a ball of radius $R_\epsilon$ around every point $p$ in a
cluster.  
A ball of radius $R_\epsilon$ centred on $p$ is usually termed the neighbourhood
of $p$. 
There are, however, two kinds of points in a cluster: 
points well inside the cluster, or {\em core points}, and points in the border,
or {\em border points}.  
A border point has less cluster points inside its neighbourhood than a core
point, as shown in left window of Fig.~\ref{theluis}. 
To include all points in a cluster, $N_{\rm MinPts}$ should be set to a low
value, what makes difficult for the algorithm to distinguish clusters from
noise. 
To solve this difficulty, DBSCAN requires that for every point $q$ in the
cluster there is another point $p$ in the cluster so that $q$ is inside the
neighbourhood of $p$, and the neighbourhood of $p$ contains at least $N_{\rm
MinPts}$ points (i.e., $p$ is a core point). 
If $q$ is in the neighbourhood of a core point but is not a core point itself
then $q$ is a border point. 
The following definition is central to turning this idea into an useful
algorithm:
a point $q$ is said {\em density-reachable} from $p$ if there is a chain of
points $p = p_1 \to p_2 \to p_3 \ldots p_n = q$ such that $p_{i+1}$ is in the
neighbourhood of $p_i$, and the neighbourhood of $p_i$ contains at least $N_{\rm
MinPts}$ points.  
A sensible definition of a cluster is, then, {\em the set of all points that are
density-reachable from a core point}, which includes all the core points as well
as all the possible border~points.  

In short, DBSCAN starts from the first point $q$ in the database and finds all
points in its neighbourhood. 
If $q$ is not a core point, DBSCAN tentatively marks $q$ as noise and proceeds
to the next point $p$. 
If $p$ is a core point, all the density-reachable points from $p$ are found and
marked as belonging to the same cluster. 
DBSCAN then proceeds to the next unclassified point in the database, repeating
the procedure until all the points are marked either as noise or belonging to a
cluster (i.e. an~agglomerate).

\subsection{Input catalogue}
\label{theinputcatalogue}

\begin{figure}
\centering
\includegraphics[width=0.48\textwidth]{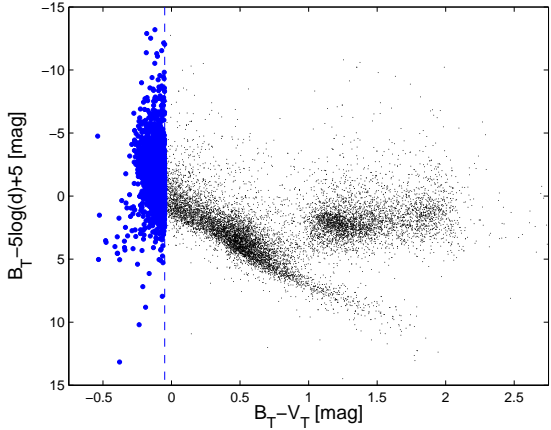}
\caption{$B_T - 5 \log{d} + 5$ vs. $B_T - V_T$ diagram of the $N_1$ = 114\,820
{\em Hipparcos} stars with no blanks in the parallax, proper motions, and $B_T
V_T$ photometry fields. 
It is roughly a colour-absolute magnitude ($M_{B_T}$) diagram, except for the
interstellar extinction factor in the $B_T$ band, that is not taken into
account. 
The $N_2$ = 4\,142 selected stars with Tycho colours $B_T-V_T <$ --0.05\,mag are
located to the left of the dashed vertical line.}
\label{thediagramblackandblue}
\end{figure}

Before looking for possible agglomerates of early-type {\em Hipparcos} stars, we
had to build the input dataset for our DBSCAN algorithm. 
By ``early-type'' we mean ``with approximate spectral types O and B''.
Slightly more than 97\,\% of the stars in the original {\em Hipparcos} catalogue
($N_0$ = 118\,218) have no blanks in the parallax, proper motions, and $B_T V_T$
photometry fields, simultaneously ($N_1$ = 114\,820). 
Among them, we selected all the sources with Tycho\footnote{For an easier
analysis, we have used the Tycho-1 magnitudes (Perryman et~al. 1997) instead of
the slightly more precise Tycho-2 ones (H{\o}g et~al.~2000).} colours $B_T-V_T
<$ --0.05\,mag ($N_2$ = 4\,142). 
No additional restriction was imposed.
These objects are the input for the clustering analysis described below.
Fig.~\ref{thediagramblackandblue} illustrates our simple colour selection
procedure. 

Except for relatively high-amplitude variability, large \linebreak 
photometric uncertainty (due to star faintness or partially resolved
multiplicity), the inaccurate Johnson-Tycho systems transformation at blue
colours, and the existence of white dwarfs and bright He-B subdwarfs below the
main sequence, the great majority of the selected sources are O- and B-type
dwarfs, giants and supergiants. 
The input dataset also contains a few stars with early-A spectral types; 
in many cases, it is due to an incorrect spectral type determination rather than
to errors in the colours. 
Of the 11 objects with Tycho colours $B_T-V_T <$ --0.05\,mag and absolute
magnitudes $M_{B_T} \approx B_T - 5 \log{d} + 5 >$ 5\,mag, six are white dwarfs
(including the intrinsically very faint, nearby, DC:-type, white dwarf
\object{GJ~440}), three are B9-type dwarfs and giants with incorrect parallax
determinations, one is a subdwarf (\object{HD~188112}), and one is an
incorrectly identified star (\object{BD+05~1825}; see
Section~\ref{possiblehipparcoserrors}).

\subsection{Parameter selection}
\label{parameterselection}

To choose the proper parameters for the analysis, we started by setting the
minimum number of bright blue stars in the cluster to a reasonable value,
$N_{\rm MinPts}$ = 6.
Decreasing this limit would lead us to consider an unmanageable amount of
agglomerates, many of which would actually be spurious groupings.

\begin{figure}
\centering
\includegraphics[width=0.48\textwidth]{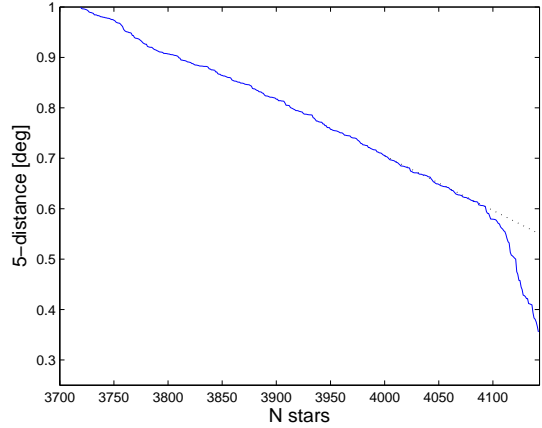}
\caption{Sorted 5-distance plot of the bright blue stars with the shortest
5th-distances in the input catalogue.   
A difference in the slope is apparent for values above and below 0.6\,deg (the
dotted line extrapolates the 5-distance for $R_\epsilon <$ 0.6\,deg).} 
\label{fig:sortedkdist}
\end{figure}

To obtain the order of magnitude of the radius $R_\epsilon$, we used the
following heuristic argument:  
first, we computed the $(N_{\rm MinPts}-1)$-distance function, mapping every
point (star) in our input catalogue with the angular distance to its fifth
nearest neighbour (``5-distance'').
We sorted the 5-distance vector by ordering all its elements in
decreasing order of their value. 
Fig.~\ref{fig:sortedkdist} shows a detail of the sorted 5-distance, that carries
information about the distribution of distances in the database (Ester et~al.
1996).
By setting $R_\epsilon$ to some fixed value, all the points (stars) with
5-distance $\le R_\epsilon$ will be core points. 
In the figure, we see that there is a critical value at $R_\epsilon \sim$
0.6\,deg that marks a change in behaviour. 
The number of points assigned to an agglomerate grows rapidly with increasing
$R_\epsilon$ below 0.6\,deg, indicating that larger increases of $R_\epsilon$
above that value are needed to see a significant change in the number of
detected agglomerates. 
Once above the critical value, the results will not vary dramatically.
From this, we can deduce that a value of $R_\epsilon \ge$ 0.6\,deg should be
used. 

On the other hand, the DBSCAN algorithm detects aggregates with density of
approximately $N_{\rm MinPts} / (\pi R_{\epsilon}^2)$ or higher.
We used a very well known cluster, the Pleiades, as a template for computing an
order of magnitude of the densities that we expect to find in our input
catalogue.
From the equality between densities, the neighbourhood radius is $R_\epsilon =
R_{\rm Ple} (N_{\rm MinPts}/N_{\rm Ple})^{1/2}$.
We have identified $N_{\rm Ple}$ = 10 classical Pleiads as {\em Hipparcos} stars
with colours $B_T - V_T <$ --0.05\,mag.  
All of them are contained within a 1\,deg-radius circle centred on Alcyone (i.e.
$R_{\rm Ple} \sim$ 1.0\,deg), \linebreak 
which makes $R_\epsilon \sim$ 0.8\,deg.
Considering this, we increased $R_\epsilon$ from $0.6$ to $0.8$\,deg in order to
decrease the threshold density of the algorithm below the density of bright blue
stars in the Pleiades.
A discussion on the effect of the \linebreak 
parametrization on the results (e.g. $R_\epsilon$ = 0.6, 0.8, 1.0\,deg) is in
Section~\ref{effectoftheparametrization}.

\section{Results}
\label{results}

   \begin{table*}
      \caption[]{Agglomerates of early-type {\em Hipparcos} stars.} 
         \label{agglomerates}
     $$ 
         \begin{tabular}{l l c c l l c l}
            \hline
            \hline
            \noalign{\smallskip}
Agglomerate 	& Reference	& $l$	& $b$	& Super- 	& Cluster/association/ 				& N$_\star$	& Possible 	\\
	  	& star		& (deg)	& (deg)	& agglomerate	& complex/region				&		& contaminant(s)\\
            \noalign{\smallskip}
            \hline
            \noalign{\smallskip}
Escorial~1 	& Alcyone	& 166.7	&--43.1	& ... 		& \object{Pleiades} 				& 10		& ...	 	\\
Escorial~2 	& V1360~Ori	& 196.1	&--22.5	& Orion		& ... 						& 6		& (Spurious agglomerate) \\
Escorial~3 	& 25~Ori	& 201.0	&--18.3 & Orion		& \object{Ori OB1~a}, \object{25 Orionis} 	& 23		& \object{HD 35716}, \object{HD 35926}~AB \\
Escorial~4    	& $\eta$~Ori	& 204.9	&--20.4	& Orion		&        {Ori OB1~a}, \object{$\eta$ Orionis}  	& 9		& \object{HD 35957} \\
Escorial~5 & $\theta^{01}$~Ori~C& 209.0	&--19.4 & Orion		& \object{Ori OB1~d}, \object{Orion Nebula}	& 23		& ...		\\
Escorial~6    	& Alnilam	& 205.2	&--17.2	& Orion		& \object{Ori OB1~b}, \object{Collinder~70} 	& 23		& ...		\\
Escorial~7    	& $\sigma$~Ori	& 206.8	&--17.4	& Orion		&        {Ori OB1~b}, \object{$\sigma$ Orionis}	& 7		& ...		\\
Escorial~8    	& 10~Mon	& 214.5	& --7.4	& ...		& \object{NGC 2232}				& 9		& \object{HD 45153}, \object{HD 45975} \\
Escorial~9    	& S~Mon~AB	& 202.9	&  +2.2	& ...		& \object{NGC 2264}, Christmas Tree		& 6		& ...		\\
Escorial~10   	& EZ~CMa	& 234.8	&--10.1	& CMa-Pup 	& \object{Collinder 121}			& 21		& ...		\\
Escorial~11   	& HH~CMa	& 233.5	& --8.7	& CMa-Pup 	&	 {Collinder 121}			& 19		& ...		\\
Escorial~12   	& 24~CMa	& 235.6	& --8.2	& CMa-Pup 	&	 {Collinder 121}			& 32		& \object{HD 53123} \\
Escorial~13   	& FV~CMa	& 236.0	& --7.3	& CMa-Pup 	&	 {Collinder 121}			& 7		& ...		\\
Escorial~14   	& HD~55019	& 240.7	& --9.0	& CMa-Pup 	&	 {Collinder 121}			& 6		& ...		\\
Escorial~15   	& 27~CMa~AB	& 239.0	& --7.1 & CMa-Pup 	&	 {Collinder 121}			& 15		& \object{$\omega$ CMa} \\
Escorial~16   	& HD~55879	& 224.7 &  +0.4	& ...		& \object{NGC 2353}				& 9		& ...		\\
Escorial~17   	& HD~56342	& 243.0	& --8.8	& CMa-Pup 	& \object{Collinder 132}			& 11		& \object{HD 56342} \\
Escorial~18   	& $\tau$~CMa~AB	& 238.2	& --5.5	& CMa-Pup 	&	 {Collinder 121}, \object{NGC~2362}	& 16		& ...		\\
Escorial~19   	& NO~CMa	& 244.9	& --7.9	& CMa-Pup 	& \object{Collinder 140}			& 10		& \object{HD 58285} \\
Escorial~20   	& HD~59138	& 241.0	& --4.7	& CMa-Pup 	&	 {Collinder 121}			& 9		& \object{HD 59364} \\
Escorial~21   	& BD--14~2020~AB& 231.0	&  +3.1	& ...		& \object{M 47}					& 6		& ...		\\
Escorial~22   	& HD~61071	& 240.4	& --2.2	& CMa-Pup 	&	 {Collinder 121}			& 6		& \object{m Pup} A(C)B (?) \\
Escorial~23   	& HD~61987	& 243.2	& --2.6	& CMa-Pup 	& ({Collinder 121})				& 8		& \object{k$^{01}$~Pup} AB, \object{HD 61687} AB \\
Escorial~24   	& c~Pup		& 252.4	& --6.7	& Pup-Vel 	& \object{NGC 2451 A}, \object{NGC 2451 B}	& 14		& \object{V468 Pup} \\
Escorial~25   	& QS~Pup	& 260.6	&--10.4	& Pup-Vel 	& \object{Vel OB2}, {\em P~Puppis}		& 10		& \object{HD 63007} \\
Escorial~26   	& $\gamma$~Vel	& 262.8	& --7.7	& Pup-Vel 	&	 {Vel OB2}, \object{NGC 2547}		& 21		& \object{HD 67704}, \object{HD 67847},	\object{HD 68657} \\
Escorial~27   	& OS~Pup	& 254.0	& --1.0	& Pup-Vel 	& \object{vdBH 23}, \object{Pup OB3}		& 11		& \object{HD 68450} \\
Escorial~28   	& NO~Vel~AB	& 262.9	& --6.9	& Pup-Vel 	&	 {Vel OB2}				& 6		& ...		\\
Escorial~29   	& IT~Vel	& 263.4	& --6.3	& Pup-Vel 	&	 {Vel OB2}				& 10		& \object{HD 70251} \\
Escorial~30   	& HD~72997	& 262.9	& --2.6	& Pup-Vel 	& \object{vdBH 34}, \object{Pismis 4}, {Vel OB2}& 6		& ...		\\
Escorial~31   	& $o$~Vel	& 270.3	& --6.8	& Pup-Vel 	& \object{IC 2391}				& 7		& ...		\\
Escorial~32   	& HX~Vel	& 266.6	& --3.6	& Pup-Vel 	& \object{IC 2395}				& 6		& \object{HD 74273} \\
Escorial~33   	& HD~75387	& 262.8	&  +0.7	& Pup-Vel 	& \object{Trumpler 10}  			& 10		& ...		\\
Escorial~34   	& HD~194670	&  78.1	&  +1.1	& ...		&       ``HIP 98321''				& 8		& ...		\\
Escorial~35   	& 8~Lac~AB	&  96.4	&--16.1	& ...		& \object{Lac OB1}				& 6		& ...		\\
            \noalign{\smallskip}
            \hline
         \end{tabular}
     $$ 
   \end{table*}
%

\begin{figure}
\centering
\includegraphics[width=0.48\textwidth]{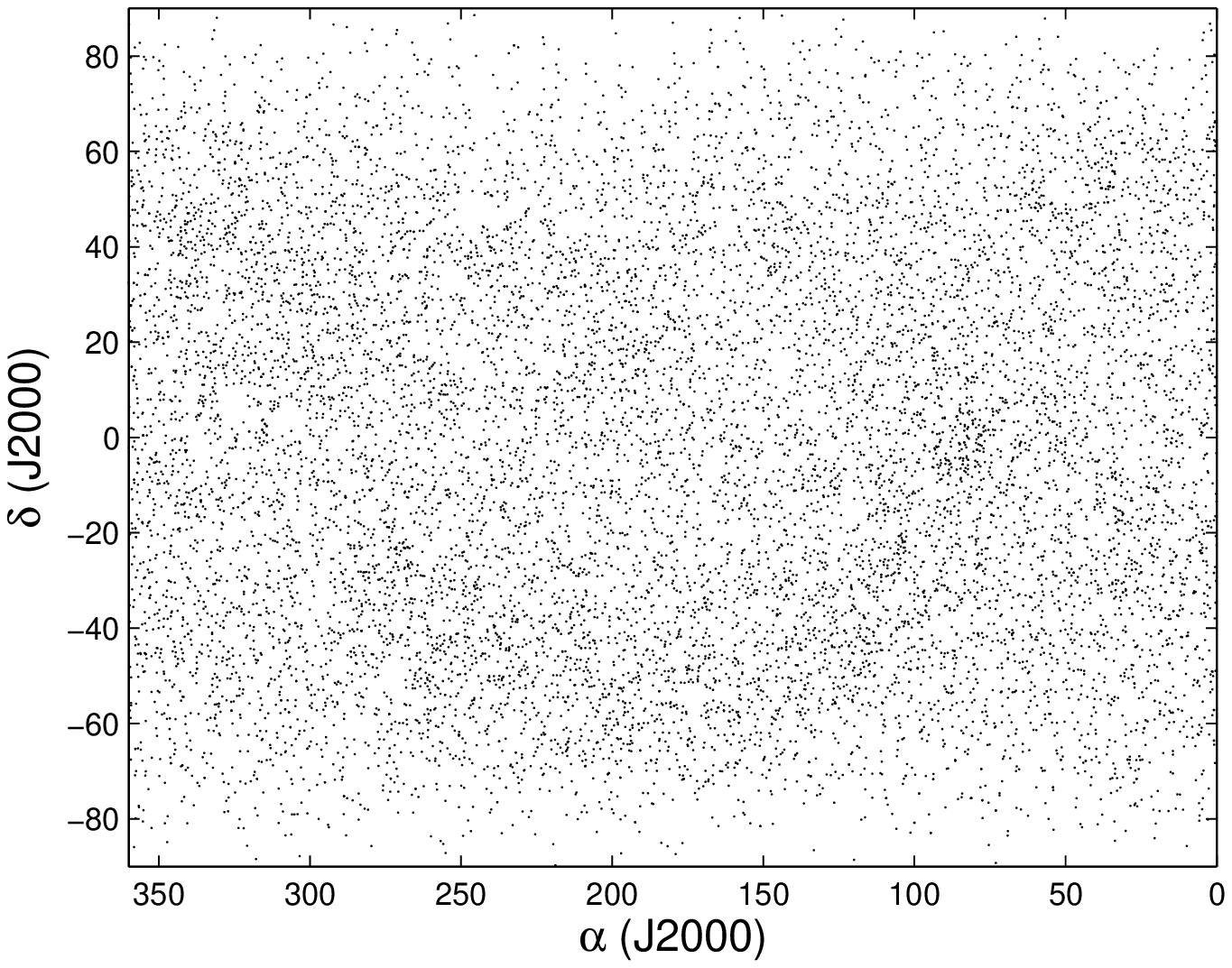}
\includegraphics[width=0.48\textwidth]{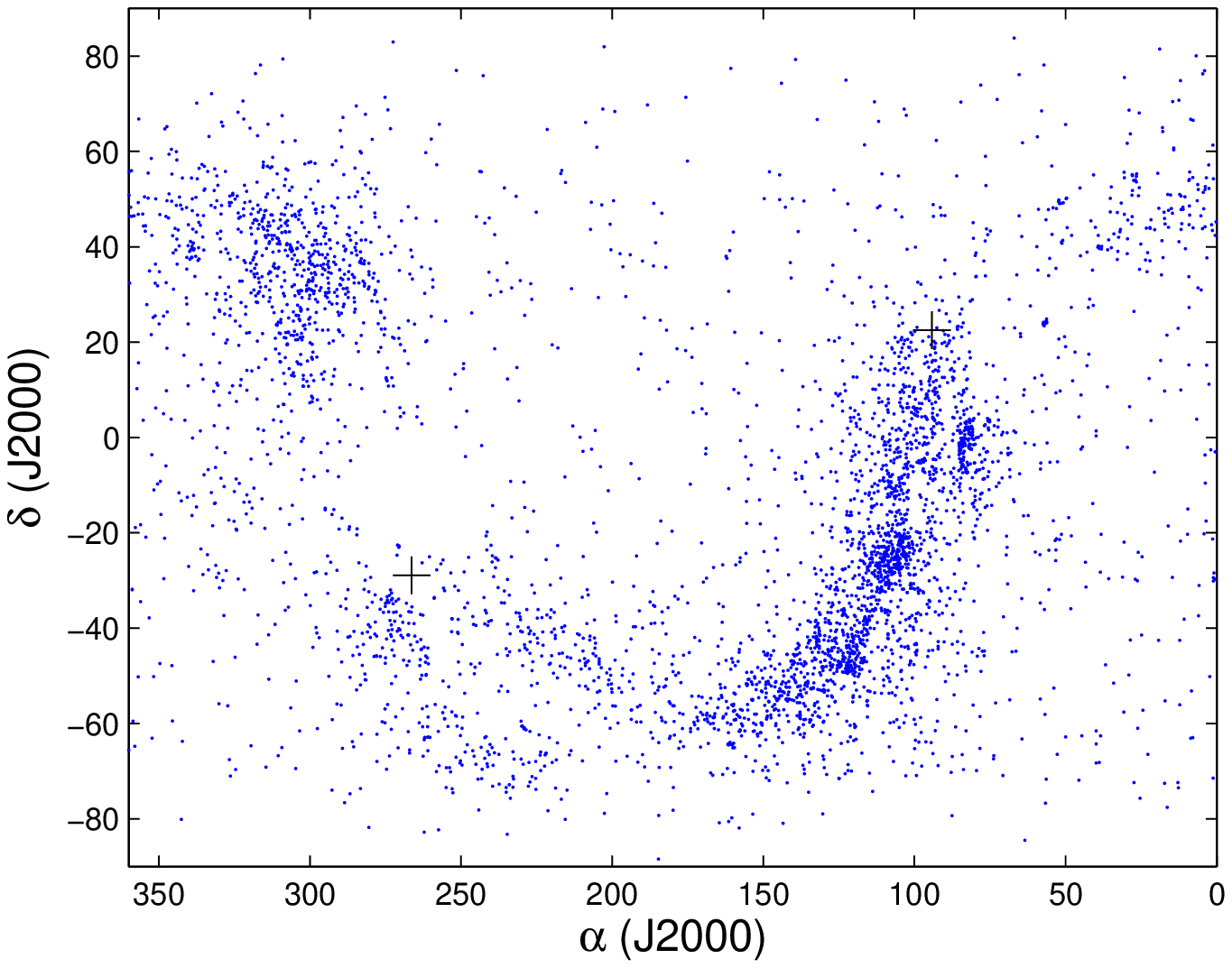}
\includegraphics[width=0.48\textwidth]{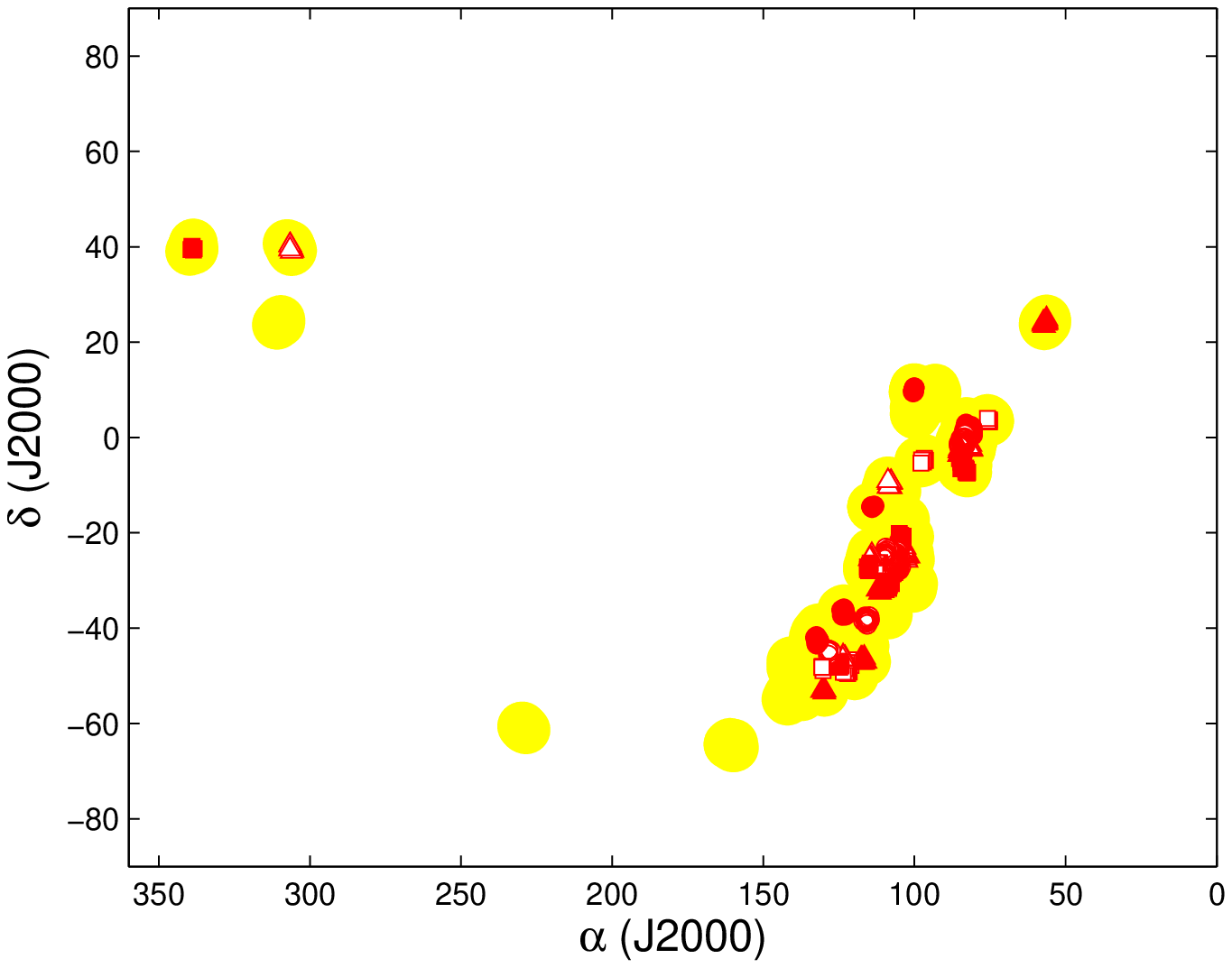}
\caption{Spatial distribution of early-type {\em Hipparcos} stars in agglomerates.
{\em Top window:} all the $N_1$ = 114\,820 {\em Hipparcos} stars with no blanks
in the parallax, proper motions, and $B_T V_T$ photometry fields. 
{\em Middle window:} $N_2$ = 4\,142 selected stars with Tycho colours $B_T-V_T
<$ --0.05\,mag.
Crosses mark the Galactic centre (left/southeast) and anticentre
(right/northwest). 
The plane of the Galactic disc is clearly discernible.
The void of blue stars to the north of the Galactic centre is the
Ophiuchus-Serpens region.
Most of the apparent overdensities are discussed next.
{\em Bottom panel:} blue {\em Hipparcos} stars in agglomerates with DBSCAN
parameters $N_{\rm MinPts}$ = 6 and $R_\epsilon$ = 0.8 (small --red-- different
symbols), and $N_{\rm MinPts}$ = 6 and and 1.0\,deg (large --yellow-- filled
circles).
Escorial~34 (open triangles) and~35 (filled squares) are the easternmost
agglomerates. 
}
\label{thegalaxy}
\end{figure}

Using the parameters $N_{\rm MinPts}$ = 6, $R_\epsilon$ = 0.8\,deg, our
DBSCAN clustering algorithm has identified 406 {\em Hipparcos} stars with
$B_T-V_T <$ --0.05\,mag distributed among 35 agglomerates.
These agglomerates are summarized in Table~\ref{agglomerates}.
We follow the nomenclature {\em Escorial}\footnote{El Escorial, a Madritian
municipality famous for its nearby Monasterio de El Escorial (a UNESCO World
Heritage Site), is the hometown of both JAC and~LD.} plus a running number to
designate the agglomerates.  
The number of stars per agglomerate varies between N$_\star$ = 6 (the minimum
number) and~32.
The spatial location of all the {\em Hipparcos} stars and those with blue
colours are shown in top and middle windows in Fig.~\ref{thegalaxy}.

Eight of the agglomerates are ``isolated'', while the remaining ones seem to be
associated in \linebreak
``super-agglomerates'' (i.e. agglomerates of agglomerates).
Although we have not carried out a quantitive study to distinguish
super-agglomerates, they naturally arise from a \linebreak
simple visual inspection of the results.
Besides, the existence of these super-agglomerates is endorsed by lots of
previous works.
One of our super-agglomerates is linked to the very famous Orion complex.
The other large agglomeration of agglomerates, that occupies an extended area in
the Canis Majoris-Puppis-Vela region (CMa-Pup-Vel; Kaltcheva \& Hilditch 2000),
has been splitted into two differentiated super-agglomerates for simplicity.
We have named them Canis Majoris-Puppis (CMa-Pup) and Puppis-Vela (Pup-Vel).
The associations Ori~OB1 (Orion), Collinder~121, \linebreak 
CMa~OB1, Pup~OB1, and Vel~OB2 (in CMa-Pup-Vel) belong to the local spiral arm of
the Galaxy (Humphreys \linebreak 
1978). 
Many of the known clusters and associations were also tabulated by Piskunov
et~al. (2006) as member candidates of the open cluster complex \object{OCC~1}
(which is apparently a signature of Gould's~Belt). 

The super-agglomerates, agglomerates and respective \linebreak 
stellar components are described next and listed in detail in
Table~\ref{theaggregates}. 
Heliocentric distances in parenthesis are from the original {\em Hipparcos}
reduction (Perryman et~al. 1997), with lower eeror bars (only in these
particular cases) than in van~Leeuwen (2007a).
Possible fore- and background blue contaminants or interlopers are also
discussed. 
Images of the agglomerates can be found in the Appendix in the electronic
version of the journal (Figs.~\ref{theaggregates.1-3}
to~\ref{theaggregates.34-35}).

\subsection{Isolated agglomerates}
\label{theisolatedagglomerates}

\begin{figure}
\centering
\includegraphics[width=0.48\textwidth]{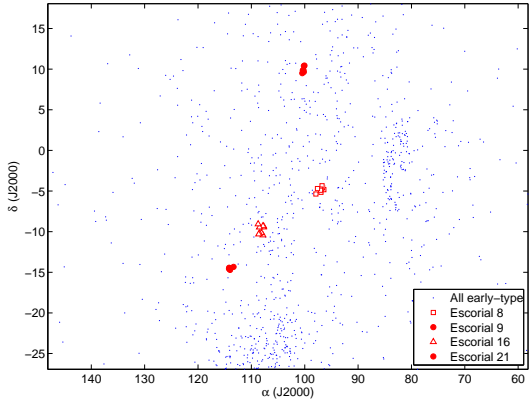}
\caption{Spatial location of the stars of the agglomerates Escorial~8 (NGC~2232)
and Escorial~9 (NGC~2264) in Monoceros, and Escorial~16 and Escorial~21
(M~17) in Puppis, in (red) different symbols (from west to east).  
Other early-type {\em Hipparcos} stars are marked with tiny (blue) dots.
Note the overdensities corresponding to the Orion complex (at $\alpha \sim$
80--85\,deg, $\delta \sim \pm$5\,deg) and the north of the CMa-Pup agglomerate
(to the south of the~plot).} 
\label{theaggregates.monocerospuppis}
\end{figure}

Our clustering algorithm has identified seven agglomerates that are not
ascribed to the Orion, CMa-Pup, or Pup-Vel super-agglomerates.
Many of them are associated to known young open clusters or star-forming
regions. 
Furthermore, a few of them, like the Pleiades (the template cluster for our
$R_\epsilon$ choice), NGC~2264, or IC~2391, are among the best studied open
clusters. 
Four of the seven agglomerates, in the Monoceros-Puppis region, are shown in
Fig~\ref{theaggregates.monocerospuppis}.

\subsubsection{Escorial~1 (Pleiades)} 
\label{escorial1}

The Pleiades (M~45, Melotte~22, the Seven Sisters) is the open cluster {\em par
excellence}. 
Known since antiquity, it is the most obvious cluster to the naked eye.
Given its short heliocentric distance, the Pleiades stars have been widely used
as calibrators for evolutionary models and as star candles (Johnson \& Mitchell
1958; Vandenberg \& Bridges 1984; Soderblom et~al. 1993; Stauffer et~al. 1994). 
The first brown dwarf, \object{Teide~1}, was discovered in this cluster (Rebolo,
Zapatero Osorio \& Mart\'{\i}n~1995).

The ten stars in the Escorial~1 agglomerate, with B6--8 spectral types and
different classes of luminosity (III, IV, and V; see Table~\ref{theaggregates}),
have magnitudes, proper motions, and parallaxes consistent with membership in
the Pleiades cluster. 
Although we have identified the ten {\em Hipparcos} stars in the region with the
bluest $B_T - V_T$ colours as members of the agglomerate, they do not match the 
classical list of the brightest cluster stars. 
Absent classical Pleiads are \object{Asterope} and \object{Celaeno}, with $B_T -
V_T \ge$ --0.05\,mag, while other identified, blue, not-so-bright Pleiads
are 18~Tau and HD~23753. 
No contaminants have been identified.
See Section~\ref{distance.pleiades} for a discussion on the Pleiades
heliocentric~distance.

\subsubsection{Escorial~8 (NGC~2232)} 

The NGC~2232 is a poorly known, young, open cluster in Monoceros.
Its brightest members is the $\beta$~Cep-type variable star 10~Mon (B2V).
Clari\'a (1972) carried out the first detailed photometric study of the cluster,
although Herschel (1864), Dreyer (1888), and Collinder (1931) had previously
catalogued it. 
There have been few late spectroscopic, photoelectric, and astrometric analyses
(Levato \& Malaroda \linebreak 
1974; Pastoriza \& Ropke 1983; Jenkner \& Maitzen 1987).
More recently, Lyra et~al. (2006) determined nuclear and contraction ages in the
narrow interval 25--32\,Ma, and an heliocentric distance of $d$ =
320$\pm$30\,pc. 
To finish the short historical review, Dias et~al. (2002) tabulated an angular
diameter of about 1\,deg. 

Our list of nine {\em Hipparcos} stars in the Escorial~8 agglomerate roughly
matches those of NGC~2232 member candidates by Robichon et~al. (1999) and
Baumgardt et~al. (2000), as well as the brightest stars listed by Clari\'a
(1972). 
The slight differences consist in the nature of a couple of possible
interlopers.
We have identified two contaminants, HD~45153 and HD~45975.
They are two B8--9V-type stars at $d \sim$ 200\,pc and proper motions
inconsistent with membership in cluster.
The NGC~2232 cluster seems to be located, from our {\em Hipparcos} data, at
330$\pm$50\,pc.
This distance agrees very well with the value derived by Lyra et~al. (2006)
from main-sequence fitting.

\subsubsection{Escorial~9 (NGC~2264)} 
\label{escorial9}

The NGC~2264 cluster in the \object{Mon~OB1} association contains four
structures that show clear evidence of extreme youth (age $\sim$ 3\,Ma) and
on-going star formation: 
the \object{Christmas Tree Cluster}, the \object{Fox Fur Nebula}, the
\object{Snowflake Cluster}, and the \object{Cone Nebula}.
Although the four structures have been considered as a single entity (precisely
called \linebreak 
NGC~2264), a purist would contradict this statement: the stellar cluster
that is actually identified in the optical is the Christmas Tree Cluster,
including the bright pre-main sequence star \object{S~Mon}~AB (O7Ve).
The latter star illuminates the nearby Fox~Fur Nebula and the more distant Cone
Nebula (the tip of which is just a few arcmin away from the pre-main sequence
giant \object{HD~47887} -- B2III:).
The nicknamed Snowflake Cluster is a dense collection of embedded, low-mass
stars detected at mid-infrared and submillimetric wavelengths, adjacent to the
\object{NGC~2264~IRS2} source (Young et~al. 2006), and at half distance between
S~Mon~AB and the Cone Nebula.   
The ``NGC~2264'' cluster as a whole has been widely used to test stellar
formation scenarios, measure frequencies of discs, and study T~Tau stars
and outflows (e.g. Walker 1956; Strom et~al. 1972; Rebull et~al. 2002; Lamm
et~al. 2004; Dahm \& Simon 2005). 

From our clustering analysis, only six blue {\em Hipparcos} stars (the minimum
number) belong to the Escorial~9 agglomerate, being S~Mon~AB the brightest one.
Of the remaing five stars, three are relatively well known pre-main-sequence
stars, and two are of unknown nature.
The latter stars, HD~47754 (B9V) and HD~47662 (B7V, a low-amplitude photometric
variable -- Koen \& Eyer 2002), are located quite to the north of the compact
H~{\sc ii} region on the centre of NGC~2264.
Their area is still rich in emission-line and variable stars, as well as
Herbig-Haro objects, so the two early-type stars may also belong to the young
star-forming region.

The close separation between the components of \linebreak 
S~Mon~AB ($\rho$ = 2.909$\pm$0.013\,arcsec, $\Delta H_p$ = 3.33$\pm$0.06) may
have affected the {\em Hipparcos} parallactic distance determination of the
system, which is clearly incorrect ($d$ = 280$\pm$40\,pc; S~Mon~AB would be
located at a significatively shorter heliocentric distance than the other
members of \linebreak 
NGC~2264, at about 600--800\,pc).

\subsubsection{Escorial~16} 

The Escorial~16 agglomerate contains nine stars.
Two of them, HD~55879 and HD~55755, were classified as members of the
moderately young NGC~2353 open cluster by Hoag et~al. (1961). 
This cluster is quite compact (about 15\,arcmin) and located near the edge of
the very young \object{CMa~OB1} association (age $\sim$ 3\,Ma -- see an
excellent review on the association in de~Zeeuw et~al. 1999). 
The most comprehensive study and review of NGC~2353 to date was carried out
by Fitzgerald, Harris \& Reed (1990).
According to them, the cluster is at the same distance as CMa~OB1 ($d$ =
1200$\pm$80\,pc), but is several times older (age $\sim$ 76\,Ma). 
They concluded, therefore, that NGC~2353 and CMa~OB1 are unrelated, and that
the brightest star in the area, HD~55879 (an early B-/late O-type giant),
does not belong to the cluster, but to CMa~OB1.
HD~55755 would remain as the only {\em Hipparcos} star member of NGC~2353.
These results are not conclusive, because Hoag et~al. (1961) assumed a
younger age of NGC~2353, rather similar to the age CMa~OB1.

There are other seven stars at up $\sim$1.3\,deg to the \linebreak 
NGC~2353 centre that belong to the Escorial~16 agglomerate.
Only one of them is remarkable: HD~55135.
It was one of the first emission-line, B-type (Be) stars to be discovered
(Merrill, Humason \& Burwell 1925).
Its strong H$\alpha$ emission might indicate that it is also a few megayears old
(unless Be stars are ``the remnants of case B mass transfer in intermediate-mass
close binaries''; Pols et~al. 1991), and then HD~55135 could even be younger
than NGC~2353. 
Of the remaining stars, HD~55117 had an {\em Hipparcos} solution flagged as
unreliable (we have taken the values for its distance and proper motions from
Falin \& Mignard [1999]).
The other four stars have been only glanced by Clari\'a (1974) or have no
references at~all.

Obviously, further spectro-photometric analyses are \linebreak 
needed to disentangle between the populations of \linebreak 
NGC~2353 and CMa~OB1, and to ascertain the actual nature of the Escorial~16
agglomerate. 
It might be a random overdensity in the disperse CMa~OB1~association.

\subsubsection{Escorial~21 (M~47)} 

The agglomerate Escorial~21 is associated to the young open cluster M~47
(NGC~2422 -- Zug 1933; Smyth \& Nandy 1962; Shobbrook 1984; Nissen 1988; Barbera
et~al. 2002; Prisinzano et~al. 2003).   
It is a $\sim$100\,Ma-old cluster at \linebreak 
$\sim$470\,pc (Rojo Arellano, Pe\~na \& Gonz\'alez 1997), with a minimum
diameter of half a degree and a very low reddening.
It has been repeatedly considered as a (more distant) Pleiades~twin.

Our six stars in the Escorial~21 agglomerate are among the brightest stars of
the cluster.
There are only two missing bright stellar systems: the (at least) quintuple
system \object{BD--14~2020} (the {\em Hipparcos} reduction probably failed
because of the intense brightness of, and closeness between, the A and B
components; $V_{A,B} \sim$ 6.5--7.0\,mag, $\rho \approx$ 7.2\,arcsec) and the
spectroscopic binary \object{KQ~Lup}~AB (an A4Ia supergiant with a quite red
$B_T-V_T$ colour). 
Our list of M~47 member stars basically coincides with that of Robichon
\linebreak  
et~al. (1999). 
The weight-averaged parallactic distance from the re-reduced {\em Hipparcos}
stars of our six early-type stars is $d$ = 450$\pm$100\,pc, that also match with
previous~estimates.

%

\subsubsection{Escorial~34 and~35} 
\label{escorial3435}

The 14 stars in the Escorial~34 (N$_\star$ = 8) and~35 (N$_\star$ = 6)
agglomerates are located far away from the previous isolated clusters and from the
Orion, CMa-Pup, and Pup-Vel super-agglomerates.
They lie in the Cygnus-Lacerta region of the Galactic plane, roughly equidistant
from the North America (\object{NGC~7000}) and Cocoon (\object{IC~5146})
nebulae. 

The stars of the Escorial~34 agglomerate are embedded in the $\gamma$~Cyg
H~{\sc ii} nebula (\object{IC~1318}), whose centre roughly coincides with the
supergiant \object{Sadr} (F8Iab:, $V \approx$ 2\,mag -- Sadr is in the
centre of Cygnus' cross).
They are close to the \object{M~29} open cluster, but there is no
indication for their membership in there.
Dolidze (1961) reported a star cluster near Sadr, Dolidze~10, but we have not
been able to identify it.
The eight stars in the agglomerate are poorly known: there is some kind of
membership information only for the four easternmost ones.
The four stars (with spectral types B8 to A0 -- see Table~\ref{theaggregates})
were classified by Platais et~al. (1998) as members of the crowded
\object{``HIP~98321''} association, that occupies an extended area around the
star HIP~98321 (\object{HD~189433}) and covers parts of Cepheus, \linebreak 
Cygnus, Lyra, and Vulpecula. 
This association extends up to 12\,deg and is located at about 300\,pc from the
Sun (Madsen, Dravins \& Lindegren 2002).
Precisely, the four \linebreak 
``HIP~98321'' stars have the shortest heliocentric distances in the Escorial~34
agglomerate ($d \sim$ 260--370\,pc). 
\object{HD~194885}, with estimated spectral type A0 and an heliocentric distance
of 260$\pm$20\,pc, may be an interloper of the ``HIP~98321'' association.
The remaining four (westernmost) stars in the agglomerate, with slightly
different proper motions and poorly determined {\em Hipparcos} distances, are
probably background early-type stars of the Orion-Cygnus spiral arm (see, e.g.,
Comer\'on et~al. [1993] for a discussion on the good visibility of the stars of
this arm from the Sun position). 
As a result, the Escorial~34 agglomerate does not seem to be a single physical
grouping of~stars.

The six stars in the Escorial~35 agglomerate are members of the \object{Lac~OB1}
association (Blaauw \& Morgan 1953; Hardie \& Seyfert 1959; Guetter 1976). 
The peculiar binary \object{HD~213918}~AB is the only {\em Hipparcos} star of
the agglomerate that is not in the list of Lac~OB1 members by de~Zeeuw et~al.
(1999), but several other works support their membership in the association
(Crawford 1961; Adelman 1968; Levato \& Abt 1976).
It is obvious that the Escorial~35 agglomerate does not represent the whole
Lac~OB1 association, but a quite small fraction (de~Zeeuw et~al. [1999]
tabulated 96 {\em Hipparcos} members of the association).
Previous claims of sub-structure within Lac~OB1 have been reported (see Lesh
[1969], de~Zeeuw et~al. [1999], and references therein for a discussion on the
Lac~OB1  ``a'' and ``b'' subgroups).
As a support for our identification, Kharchenko et~al. (2005) listed four of
our {\em Hipparcos} stars (actually six stars in three multiple systems,
including the resolved binary \object{8~Lac~A} and~B) as members of the new
Galactic open cluster \object{[KPR2005]~122} (ASCC~122).
Additional studies are needed to confirm if this overdensity of early-type
stars within Lac~OB1 is a physical grouping or~not.

\subsection{The Orion super-agglomerate}
\label{theorionsuperagglomerate}

Apart from Cep~OB1, that was not so well investigated at that moment, the
Ori~OB1 complex was the most massive OB association in the 1\,kpc sphere centred
on the Sun in the classical review by Blaauw (1964).
It was, besides, more compact than the other massive Scorpius-Centaurus and
\linebreak 
Lac~OB1 complexes. 
The total mass and largest overall projected diameter estimated by him for the
Ori~OB1 complex, at about 7600\,$M_\odot$ and 100\,pc, roughly match
current determinations. 
The currently accepted division of Ori~OB1 into four subgroups was also
presented by Blaauw (1964). 
He proposed the following sub-associations:
\begin{itemize}
\item Ori~OB1a: north and west of the Orion Belt,
\item Ori~OB1b: the Orion Belt itself,
\item Ori~OB1c: south and east of the Orion Belt, except the Orion Sword,
\item Ori~OB1d: in and close to the Orion Sword.
\end{itemize}
Warren \& Hesser (1977) splitted, in its turn, the Ori~OB1b subgroup into three
sub-subgroups (b1, b2, and b3), and the Ori~OB1c subgroup into five
sub-subgroups (c, c1, c2, c3, and c4).
The nine groupings are supposed to have different ages and heliocentric
distances, in the intervals 1--10\,Ma and 350--500\,pc, respectively, being
Ori~OB1d (including the Orion Nebula Cluster and the Trapezium) the youngest
one. 
Further classical reviews on the Orion molecular cloud and star-forming region,
emphasizing on its importance in modern Astronomy, can be found at Genzel \&
Stutzki (1989) and Brown, de Geus \& de~Zeeuw (1994). 

Our DBSCAN algorithm has identified six agglomerates in the region 05 00
$\le \alpha \le$ 05 40, --8 $\le \delta \le$ +4, quite close
to the Solar antapex.
Any blue star with low proper motion in this area is capable of belonging to the
Ori~OB1 complex. 
We consider that all but one of these agglomerations are actual members of the
complex.
Since they do not exactly correspond to the classical grouping in the cluster,
we will follow the ``Escorial'' designation.
See Section~\ref{sub-structure} for a discussion on the sub-structure of the
Orion super-agglomerate.

\begin{figure}
\centering
\includegraphics[width=0.48\textwidth]{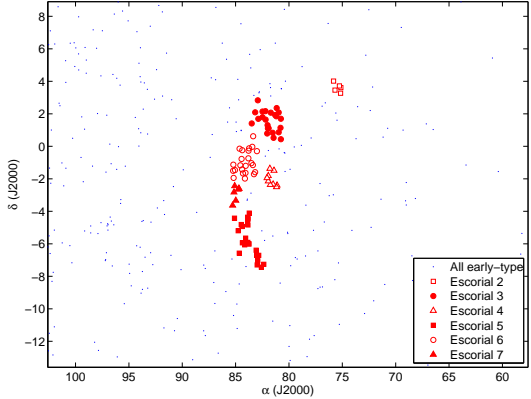}
\caption{Same as Fig.~\ref{theaggregates.monocerospuppis}, but for the
agglomerates Escorial 2--7 in the Orion super-agglomerate.}
\label{theaggregates.orion}
\end{figure}

\subsubsection{Escorial~2} 

This agglomerate of only six stars, with B8--9V spectral types, is in a
distinct location with respect to the other five agglomerates.
Far from presenting it as a new subgroup in the Ori~OB1 complex, we consider
Escorial~2 a spurious agglomeration.
This hypothesis is supported, apart from the different location, by: 
($i$) the low number of stars in the agglomerate (the minimum $N_{\rm MinPts}$), 
($ii$) their relatively late spectral types,
($iii$) the heliocentric distances shorter than 300\,pc of at least two
(possibly four) stars, and
($iv$) the presence of two of the stars in a close binary (HD~32039, HD~32040;
if they were a single star, then $N_\star$ = 5 $< N_{\rm MinPts}$ and DBSCAN
would have not recognize it as an agglomerate).

\subsubsection{Escorial~3 (25~Orionis)} 

The 23 stars in the Escorial~3 agglomerate lay on the northern part of the
Ori~OB1a subgroup, surrounding the bright B1Vpe star 25~Ori (also known as
$\psi^{01}$~Ori)\footnote{When available, we follow the Bayer (1603)
nomenclature instead of the Flamsteed (1712) one.
However, in this case there can be a misunderstanting between 25~Ori
($\psi^{01}$~Ori) and the $\beta$~Cep-variable star $\psi$~Ori (30~Ori,
$\psi^{02}$~Ori).}. 
In spite of the large amount of stars in the agglomerate, the group was not
identified until very recently, when Brice\~no et~al. (2005) and Kharchenko
et~al. (2005) discovered what they called the ``25~Ori group'' and the
``ASCC~16 cluster'' (\object{[KPR2005]~16}), respectively.
Many of the stars in our agglomerate, including 25~Ori itself, have been
catalogued in those works, \linebreak 
which supports a correct identification.
While for Brice\~no et~al. (2005) the Escorial~3 agglomerate stood out as a
concentration of T~Tau stars in the Ori~OB1a sub-association, Kharchenko et~al.
(2005) found it based on parallaxes, \linebreak 
proper motions, and $B_T V_T$ photometry for {\em Hipparcos} and Tycho-2 stars.  
The confirmation of Escorial~3 having distinct kinematics (in $V_r$) and age
(7--10\,Ma) in Ori~OB1a was presented by Brice\~no et~al. (2007).

Not all the 23 stars belong to the young agglomerate.
In Table~\ref{agglomerates}, we list two possible interloper stars with
heliocentric distances shorter than 200\,pc.
Of them, HD~35926~AB is a binary system separated by 1.3\,arcsec (Dommanget \&
Nys 1994).
Additionally, there are other three stars with $d + \delta d <$ 300\,pc
(including errorbars) that we do not classify as possible contaminants.
If we fully trust {\em Hipparcos}, \object{HD~36429} would be a B5V star at only
$d$ =  200$\pm$30\,pc.  
The presence of a close companion might have affected its parallax
measurement (see, however, Section~\ref{fielddwarfs200pc})\footnote{HD~36429 has
a previously unreported companion ($\rho \sim$ 9.7\,arcsec, $\theta \sim$
163\,deg, $\Delta K_{\rm s}$ = 3.42$\pm$0.05\,mag), apart from a known visual
companion HD~287931 ($\rho \sim$ 1.1\,arcmin, $\theta \sim$ 350\,deg, $\Delta
K_{\rm s}$ = 0.09$\pm$0.03\,mag).}. 
Although Brice\~no et~al. (2007) carried out a challenging spectroscopic study
of members in Escorial~3 (25~Ori), further analyses in the agglomerate are
desirable to disentangle the young star population from the fore-/background and
from Ori~OB1a and Ori~OB1b.

\subsubsection{Escorial~4 ($\eta$~Orionis)} 
\label{escorial4}

The nine stars in the Escorial~4 agglomerate are located in the southern part of
the Ori~OB1a subgroup, to the southwest of the Orion Belt.
All of them except HD~35957, with a parallactic distance of $d$ =
190$\pm$40\,pc, seem to be members of the Ori~OB1a subgroup based on early
spectral types, parallaxes, low proper motions and positions in the $V_T$ vs.
$B_T-V_T$ colour-magnitude diagram (see also Sharpless \linebreak 
[1952] and Warren \& Hesser [1978]).
To our knowledge, none of the stars have been previously reported to belong to
any sub-structure within Ori~OB1a.
Since the brightest star in the agglomerate is $\eta$~Ori~AB (28~Ori), we
propose the name ``$\eta$~Orionis'' for the overdensity. 
Approximate central coordinates and diameter of the overdensity, whose
hypothetical cluster nature must be confirmed, are 05~26~30 --01~55~00 (J2000)
and 1.2\,deg, respectively. 
From the comparison of cluster photometric sequences, the age of the stars in
the $\eta$~Orionis overdensity must not differ very much from those of the
25~Ori agglomerate and the  Orion Belt (e.g. 5--10\,Ma).

Brice\~no et~al. (2005) derived the basic properties ($T_{\rm eff}$, $L$, $M$)
of the seven single stars in the agglomerate, including HD~35957.
Their masses ranged in the interval 2.8--9.4\,$M_\odot$.
The remaining two stars are in close binary systems, and so it is difficult to
derive basic properties from photometry and low-resolution spectroscopy
($\eta$~Ori~AB, $\rho$ = 1.695\,arcsec; HD~35456~AB, $\rho$ = 0.810\,arcsec).

See Section~\ref{etaOri} for a discussion on the possible cluster nature of the
$\eta$~Orionis overdensity.

\subsubsection{Escorial~5 (Orion Sword)} 
\label{escorial5}

The region of the Orion Sword, including the Trapezium and the Orion Nebula
Cluster, is one of the most investigated and scrutinized areas in the 
sky, and almost does not require any introduction (Kwan 1977; Prosser et~al.
1994; Hillenbrand 1997; Lada et~al. 2000; Feigelson et~al. 2002).  
It is usually presented as ``the closest region of massive star formation to the
Sun'' (although some authors give this honour to the Scorpius-Centaurus
association).
The four stars that constitute the Trapezium, the asterism that illuminates the
Orion Nebula (M~42), are the brightest stars of the Orion Nebula Cluster. 
This cluster displays the largest stellar density in Orion (together with
$\sigma$~Orionis; see below), and is considered as a distinct subgroup,
Ori~OB1d.
Also in the Orion Sword, and running from north to south, there are other
subgroups of extremely young stars that have been classified as members of the
Ori~OB1c1--4 sub-subgroups (Warren \& Hesser 1977). 
They are differentiated overdensities in the Ori~OB1c subgroup.
The Orion Nebula Cluster is sandwiched between Ori~OB1c3 and Ori~OB1c4
sub-subgroups.

Practically all the 23 stars in the Escorial~5 agglomerate have been classified in
the literature as members of the Ori~OB1c[1--4] or Ori~OB1d subgroups (e.g.
Parenago 1954; Warren \& Hesser 1978).
Although it is obvious that the Orion Nebula Cluster has very particular
properties (e.g. very high extinction and highly concentrated radial
distribution of stars), we have not been able to disentangle it from the
remaining stellar population in the Orion Sword (i.e. the Ori~OB1c[1--4]
sub-subgroups). 
Indeed, $\theta^{01}$~Ori and the Trapezium stars ({\em Hipparcos} only
identifies three components of $\theta^{02}$~Ori) have Tycho-1 $B_T-V_T$ colours
redder than --0.05\,mag and were not, therefore, in the input catalogue for our
DBSCAN algorithm\footnote{The Tycho-2 $B_T-V_T$ colour of the hierarchical
triple $\theta^{01}$~Ori is, however, bluer than --0.05\,mag.}. 
The brightest and bluest stars in the Escorial~5 agglomerate are $\iota$~Ori~AC
(O9III+, 3\,mag-$V$ star in the southern border of M~42), $c$~Ori~AB (B1V+, in
the centre NGC~1977 nebula, to the north of M~42), HD~36960 (B0.5V, close to
$\iota$~Ori~AC), and $\upsilon$~Ori (B0V, at half a degree to the south of the
Trapezium).

\subsubsection{Escorial~6 (Orion Belt)} 
\label{escorial6}

Although the three bright O-type supergiants in the Orion Belt (Alnitak, Alnilam
and Mintaka) constitute a prominent asterism in the most obvious constellation
in the sky, it was not until late 1920s when Pannekoek (1929) noticed a
``clustering of early-type stars elongated roughly parallel to the galactic
plane''.
The three supergiants and dozens young stars in the Orion Belt, {\em including}
$\sigma$~Orionis, have been considered since Blaauw (1964)'s work to belong to
the Ori~OB1b subgroup. 
A complete review of the (sub-)stellar populations around the three supergiants
is provided in Caballero \& Solano (2008).
The latter authors describe the Orion Belt as the combination of:
($i$) a highly extinguished star-forming region surrounding Alnitak
($\zeta$~Ori),  
($ii$) a wide, populated, low-density cluster with no clear central overdensity
surrounding Alnilam ($\epsilon$~Ori), that is spatially coincident with the
cluster Collinder~70 (Collinder 1931), and
($iii$) a recently-discovered cluster candidate surrounding Mintaka
($\delta$~Ori). 

The 23 stars in the Escorial~6 agglomerate have been previously considered
members of the Ori~OB1b subgroup (see, again, Warren \& Hesser 1978).
Our list (Table~\ref{theaggregates}) contains the three supergiants and other
well known early-type stars and multiple systems, like the eclipsing
spectroscopic binary \object{VV~Ori}~AB, the likely Lindroos system\footnote{A
Lindroos system is a binary with a main sequence, early-type primary and a
post-T~Tau secondary.} \linebreak 
\object{HD~36779}~AC--B, or the helium strong, chemically peculiar, variable
star \object{V901~Ori}.  
No possible contaminants have been identified.
The blue {\em Hipparcos} stars in the vicinity of the Horsehead Nebula and the
$\sigma$~Orionis cluster are {\em not} contained in the Escorial~6 agglomerate
(see~below).

\subsubsection{Escorial~7 (Horsehead)} 

The Escorial~7 agglomerate contains seven stars, four of which were listed by
Caballero (2007) among the brightest stars of the $\sigma$~Orionis cluster.
Both $\sigma$~Ori~AF--B and D (together with the other components of the
multiple \linebreak 
Trapezium-like system, not catalogued by {\em Hipparcos}; Caballero 2008b) are
in the centre of the eponym cluster. 
\linebreak 
HD~37525~AB, at 5\,arcmin to the southeast of the centre, is still in the
``core'' of the cluster ($r \le$ 20\,arcmin; Caballero 2008a).
The last bright cluster member, HD~37699, lies on the ``halo'' of the cluster,
where the contamination by (overlapping) neighbouring young star populations may
be large (e.g. Jeffries et~al. 2006b).
The B1.5V variable star HD~37744, at a bit more than 30\,arcmin from the cluster
centre, has occasionally been considered as a $\sigma$~Orionis \linebreak 
member (Sherry, Walter \& Wolk 2004), but it has not been listed in the recent
Mayrit catalogue of stars and brown dwarfs in the cluster (Caballero 2008c).
The other (chemically peculiar) two stars, V1148~Ori and HD~37807, are much
farther away to the south.
The three latter stars \linebreak 
(HD~37744, V1148~Ori, and HD~37807) are, besides, spatially coincident with the
north-to-south arrangement of the IC~434/NGC~2023 complex, whose more patent
structure is the Horsehead Nebula. 
Because of this coincidence, and of $\sigma$~Ori illuminating the mane of the
Horsehead, we call the Escorial~7 agglomerate ``the Horsehead region''.
Increasing from $R_\epsilon$ = 0.8 to 1.0 in our DBSCAN analysis would lead the
Escorial~6 (Orion Belt) and~7 (Horsehead) agglomerates to fuse into one
subgroup, as commonly considered. 
In any case, ``the Horsehead region'' seems to be the juxtaposition of a highly
concentrated cluster ($\sigma$~Orionis), to the northwest, and a younger
population of stars associated to the Horsehead Nebula, to the east and
southeast.

\subsection{The CMa-Pup super-agglomerate}
\label{thecmapupsuperagglomerate}

\begin{figure}
\centering
\includegraphics[width=0.48\textwidth]{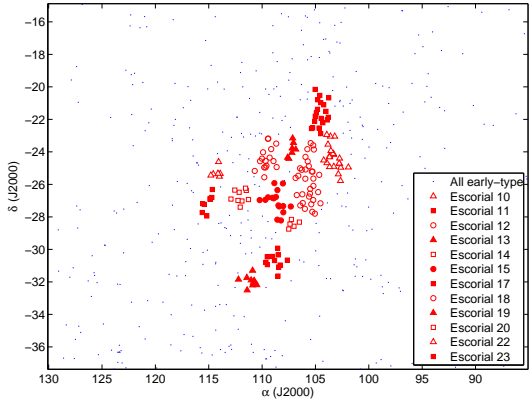}
\caption{Same as Fig.~\ref{theaggregates.monocerospuppis}, but for the
agglomerates Escorial 10--15, 17--20, 22, and~23 in the Canis Majoris-Puppis
super-agglomerate.
The two southern agglomerates (Escorial~17 and~19), at $\delta \le$
--30\,deg, belong to the Lower~CMa-Pup grouping.
The remaining agglomerates belong to the Upper~CMa-Pup grouping (the
Collinder~121 association).} 
\label{theaggregates.lcmapup}
\end{figure}

The CMa-Pup super-agglomerate is located to the north of the Canis
Majoris-Puppis-Vela region. 
For an easier \linebreak 
description, we have splitted the agglomerates in CMa-Pup into two
differentiated groupings: ``Upper~CMa-Pup'' (to the north) and ``Lower~CMa-Pup''
(to the south; Fig.~\ref{theaggregates.lcmapup}).

\subsubsection{Upper~CMa-Pup (Collinder 121)} 

The Upper~CMa-Pup grouping is formed by the agglomerates Escorial~10--15, 18,
20, 22, and~23. 
Most of the stars in the agglomerates in Upper~CMa-Pup except the
easternmost Escorial~23 one have been classified as members of the Collinder~121
cluster/association (de~Zeeuw et~al. 1999; Robichon et~al. 1999). 
According to Kaltcheva \& Makarov (2007), Collinder~121 would be the
superposition of the genuine compact group discovered by Collinder (1931), 
\linebreak 
which would be at an heliocentric distance of $\sim$1\,kpc, and an extended,
loose OB association with a large depth probably extending from $\sim$0.5 to
beyond 1\,kpc.
Escorial~10, the westernmost agglomerate in the grouping, would be the 
\linebreak 
genuine Collinder~121 cluster (Feinstein~1967).
 
In total, ten agglomerates are associated to \linebreak 
Collinder~121.
Given its proximity to the remaining agglomerates, there is no reason to exclude
the stars in Escorial~23 from the grouping. 
This is the first time in the literature to propose such a drastical division of
Collinder~121.
The extension of the region, of up to 12\,deg in both vertical and horizontal
directions, would lead to maximum physical separations between stellar
components of 100--200\,pc.
This size is not far larger than that estimated for the Orion complex
(Section~\ref{theorionsuperagglomerate}), which we have splitted into 5+1
agglomerates.
Upper~CMa-Pup/Collinder~121 might be an analog to the Orion complex, but at a
triple heliocentric distance and splitted into ten agglomerates.
The grouping contains the most populated agglomerate, Escorial~12
(N$_\star$ = 32, roughly centred on the blue supergiant
24~CMa)\footnote{Star 24~CMa is $o^{02}$~CMa; do not mistake with 16~CMa
($o^{01}$~CMa, K2.5Iab:).}.
There are also some Upper~CMa-Pup stars in known open clusters that could be
associated to Collinder~121, or be in the background (e.g. NGC~2362
surrounding \object{$\tau$~CMa~AB} in Escorial~18 -- age = 5$^{+1}_{-2}$\,Ma,
$d$ = 1480\,pc; Moitinho et~al. 2001).
Some stars are associated to reflection nebulae (e.g. \linebreak 
\object{HD~61071} in Escorial~22; van~den~Bergh~1966).

In Table~\ref{theaggregates}, we list four possible contaminants in the
foreground of Upper~CMa-Pup/Collinder~121. 
All of them have parallaxes and/or proper motions inconsistent with \linebreak 
membership in the association (the {\em Hipparcos} \linebreak 
measurements of m~Pup~AB could be affected by close binarity -- actually, it
could be a hierarchical triple system, since the primary is thought to be a
eclipsing binary; Stift 1979). 
It is interesting to notice the very early spectral types of some of these
interlopers, like the well-known, emission-line, variable star $\omega$~CMa
(B2IV/Ve -- Baade 1982; Sletteback 1982; {\v S}tefl et~al. 2003), which is
located at only $d$ = 279$\pm$13\,pc according to van~Leeuwen~(2007a).

\subsubsection{Lower~CMa-Pup: Escorial~17 (Collinder~132) and~19
(Collinder~140)}  
\label{escorial1719}

There remain two agglomerates in the CMa-Pup \linebreak 
super-agglomerate: Escorial~17 and~19. 
They are associated to Collinder~132 and~140, respectively, and form the
Lower~CMa-Pup grouping.
Each of the agglomerates has only one possible foreground contaminant
(Table~\ref{theaggregates}).

There has been only one dedicated work in \linebreak 
Collinder~132, carried out by Clari\'a (1977), and there still remain caveats on
the actual nature and structure of the cluster. 
First, Clari\'a (1977) interpreted that there are two separate physical groups,
called ``Cr~132a'' and ``Cr~132b'', located at 560 and 330\,pc from the Sun, and
with nuclear ages of 60 and 160\,Ma, respectively. 
Eggen (1982) and Baumgardt (1998) supported the double cluster hypothesis, but
with nuance. 
Baumgardt (1998) stated that ``Collinder 132 is found to be mainly composed out
of members of an OB association, but there may be a star cluster present in this
area too''.  
He also proposed that there may be a connection between Collinder~132 and
Upper~CMa-Pup/Collinder~121. 
However, Robichon et~al. (1999) derived a unique parallactic distance of $d$ =
650$\pm$140\,pc, much further away than the 270\,pc tabulated by Wielen (1971),
who took the value, in its turn, from Collinder (1931).
Other authors have catalogued closer distances ($d \approx$ 400\,pc) and younger
ages ($\sim$25\,Ma; e.g. Battinelli, Brandimarti \& \linebreak 
Capuzzo-Dolcetta~1994).

Our Escorial~17 agglomerate possesses 11 stars.
The Collinder~132 cluster centre has been considered to fall \linebreak 
close to the brightest star in our agglomerate, HD~56342 (B3V).
Indeed, this star was classified as the brightest cluster member by Clari\'a
(1977). 
HD~56342 has, however, parallax and proper motion measurements clearly different
\linebreak 
from the rest of the stars of the agglomerate.
This difference led us to classify it as a young contaminant star in the
foreground. 
Obviously, an isolated bright B3V dwarf at only $d$ = 193$\pm$8\,pc is very
uncommon;
it will be discussed in Section~\ref{fielddwarfs200pc}. 
From our data, we identify only one structure with heliocentric distance and
age consistent with those provided by Robichon et~al. (1999) ($d$ =
650$\pm$140\,pc) and Battinelli et~al. (1994) (age $\sim$ 25\,Ma).
There might be a cluster of {\em faint} stars in the foreground, associated to
HD~56342, but we fail to identify~it.

Collinder~140, although discovered by Collinder (1931), could have been
previously identified by de Lacaille (1755), who gave it the name ``Nebulous
Star Cluster No.~II.2'' ({\em aka} Lacaille~II.2).
Since Collinder~140 contains at least five stars brighter than $V$ =
7\,mag in a concentrated arrangement, the cluster has received a larger
attention than Collinder~131. 
Williams (1967a, 1978), Fitzgerald, Harris \& Miller (1980), Lyng{\aa} \&
Wramdemark (1984), and other authors have investigated Collinder~140 in detail.
Williams (1967b) suggested that Collinder~140 and NGC~2451
(Section~\ref{escorial2427}), together with other two open clusters, are ``the
remaining nuclei of an OB association'' that broke~up. 
Clari\'a \& Rosenzweig (1978) carried out the most complete investigation in
Collinder~140.
They derived its heliocentric distance ($d$ = 365$\pm$29\,pc), nuclear and
contraction ages (at about 20--25\,Ma), physical size ($\sim$10\,pc), total mass
($\ge$ 100\,$M_\odot$), \linebreak 
number of evolved members ($\ge$ 3), metallicity ([Fe/H] = --0.1), and other
parameters, like the cluster radial velocity or the Galactic space motion. 
Different heliocentric distances and ages have been provided afterwards
(Williams 1978: $d$ = 420$\pm$20\,pc, age $\sim$ 40\,Ma;
Fitzgerald et~al. 1980: $d$ = 410$\pm$30\,pc, age = 20$\pm$6\,Ma --they pointed
out a significant concentration of yellow giants--;
Robichon et~al. 1999: $d$ = 410$^{+60}_{-40}$\,pc). 

One of the brightest stars in Collinder~140 is the variable supergiant HD~58535
(G8II, $V$ = 5.35\,mag -- Harris 1976; Clari\'a 1976).
It has a very red $B_T-V_T$ colour and was not in the input catalogue of our
DBSCAN algorithm.
The rest of the {\em Hipparcos} stars in Escorial~19 have also been classified
as members of Collinder~140, except for the binary HD~59499+HD~59500 (B3V+B4V),
that lies at about 1\,deg to the east of the cluster centre. 
We classify the binary as a Collinder~140 candidate member for the first time.
Robichon et~al. (1999) listed other three {\em Hipparcos} stars with $B_T-V_T >$
--0.05\,mag that are not, therefore, in our compilation.
Besides, we have classified the EA-type eclipsing binary HD~58285 (B9III, $P$ =
2.198515\,d, in eccentric system -- Houk 1982; Otero \& Dubovsky 2004) as a
possible contaminant in the outer part of Collinder~140, because of its
distinguishable proper motion. 
The weighted mean heliocentric distance to Collinder~140, from our {\em
Hipparcos} data, is $\overline{d}$ = 370$\pm$50\,pc.
It is similar to the distance measured by Clari\'a \& Rosenzweig (1978), and
consistent within errorbars with other recent determinations.

\subsection{The Pup-Vel super-agglomerate}
\label{thepupvelsuperagglomerate}

The Pup-Vel super-agglomerate is, as well as CMa-Pup, in the Canis
Majoris-Puppis-Vela region. 
However, although our DBSCAN algorithm has also splitted it into ten
agglomerates, the Pup-Vel super-agglomerate is not as compact as the CMa-Pup
one. 
The ten agglomerates in Pup-Vel are distributed in a hierarchical spatial
distribution. 
The largest spatial density coincides with a dense part of the Vel~OB2
association (Section~\ref{escorial262829}). 
Surrounding this ``core'', there are other seven well-defined clusters, some of
which have been repeteadly investigated in the literature (IC~2391, \linebreak 
Trumpler~10, and NGC~2451~AB). 

\begin{figure}
\centering
\includegraphics[width=0.48\textwidth]{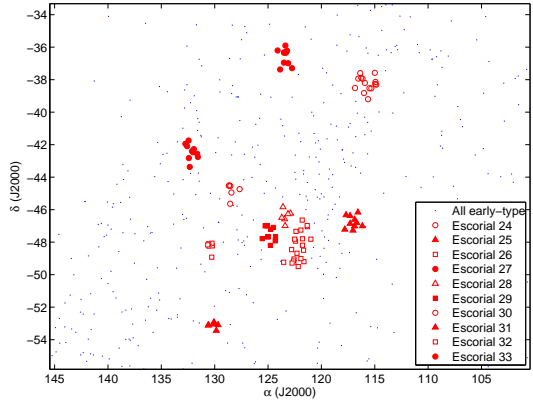}
\caption{Same as Fig.~\ref{theaggregates.monocerospuppis}, but for the
agglomerates Escorial 24--33 in the Puppis-Vela super-agglomerate.
The trio Escorial 26--28 is in the centre; Escorial~30, IC~2391 (Escorial~31),
IC~2395 (Escorial~32), and Trumpler~10 (Escorial~33) are to the east; 
NGC~2451~AB (Escorial~24) and vdBH~23 (Escorial~27) are to the north;
and Escorial~25 is to the~west.}
\label{theaggregates.puppisvela}
\end{figure}

\subsubsection{Escorial~26, 28, and~29} 
\label{escorial262829}

Escorial~26, 28, and~29 form a dense trio of agglomerates, usually ascribed to
the Vel~OB2 association. 
The largest and most populated agglomerate is Escorial~26, that has an elongated
shape in the north-south direction.
Although our DBSCAN algorithm has not been able to disentangle both populations,
the Escorial~26 agglomerate is actually the juxtaposition of the
$\gamma$~Velorum star-forming region, to the north, and the NGC~2547 cluster, to
the south.  The open cluster Collinder~173 is at the northern borderline of the
agglomerate. 

On the one hand, the $\gamma$~Velorum region practically coincides with the
densest part of the Vel~OB2 association, that was identified by Kapteyn (1914)
and described in detail by de~Zeeuw et~al. (1999)\footnote{Do not mistake with
Eggen (1982)'s Vel~OB2, that corresponds to another group of~stars.}.
The (at least) sextuple $\gamma$~Vel star ($\gamma^{02}$~Vel) is in the centre
of the association, and is one of the brightest stars in the nighttime sky ($V
\approx$ 1.8\,mag).
The brightest component of $\gamma$~Vel is actually a spectroscopic binary
composed of WR~11 (WC8), the nearest known Wolf-Rayet star, and a massive O9I
supergiant with a 78.5-day orbital period.
An heliocentric distance of 411$\pm$12\,pc was derived by de~Zeeuw et~al.
(1999).
They also listed literature ages in the interval 6--20\,Ma.
Interestingly, the \object{Vela~Pulsar} was born in Vel~OB2 about 10$^4$\,a ago
(Hoogerwerf, de Bruijne \& de~Zeeuw 2001).
On the other hand, NGC~2547 has been very well investigated.
The cluster is very important for studying X-ray emission (Jeffries \& Tolley
1998; Jeffries et~al. 2006a), the Wielen dip\footnote{The Wielen dip is a
feature in the luminosity function at $M_V \approx$ 7--8\,mag. 
NGC~2547 is the youngest open cluster to display it.} (Naylor et~al. 2002), the
lithium depletion boundary (Jeffries \& Oliveira 2005), the disc frequency
(Young et~al. 2004; Gorlova et~al. 2007), low-mass stars and brown dwarfs
(Jeffries et~al. 2004), mass segregation (Littlefair et~al. 2003), and rotation
and activity (Jeffries, Totten \& James 2000; Irwin et~al. 2008). 
It is 30$\pm$10\,Ma old (see Oliveira et~al. [2003] for a discussion on the
discrepancy between ages from isochrone fitting and from the lithium depletion
boundary) and at 400--450\,pc, and has a low reddening (Fernie 1959, 1960;
Clari\'a 1982 and references above). 
In Table~\ref{agglomerates}, we list three possible contaminants in the
Escorial~26 agglomertae (HD~68657, at $d$ = 259$\pm$18\,pc and with a distinct
proper motion, has an early spectral type B3V -- Hiltner, Garrison \& Schild
1969--. 
It might be a B-type subdwarf in the foreground).

The other 16 stars in the trio are located to the east and northeast of
Escorial~26, overlapping with the Vela supernova remnant.
They are distributed among two agglomerates: Escorial~28 (northeast, $N_\star$ =
6) and Escorial~29 (east, $N_\star$ = 10).
HD~70251 could be a foreground interloper in the latter agglomerate (it is a B8V
stars at $d$ = 290$\pm$30\,pc and discordant proper motion).
Most of the remaining stars have been classified as members of the Vel~OB2
association.
Remarkably, \object{HD~70309~A} (B3III) is the primary of a Lindroos system (the
secondary, \object{HD~70309~B}, is a K2IV star -- Lindroos 1985; Pallavicini,
Pasquini \& Randich~1992).

To sum up, we find evidence of subclustering in \linebreak 
Vel~OB2.
Besides, the similarity in heliocentric distance and age between Vel~OB2,
$\gamma$~Velorum, and NGC~2547 may suggest that they actually belong to the same
complex, but \linebreak 
NGC~2547 having a much larger spatial density of members.

\subsubsection{Escorial~30, 31 (IC~2391), 32 (IC~2395), and~33 (Trumpler~10)} 
\label{escorial30313233}

The four agglomerates Escorial~30--33 are located to the east of the Pup-Vel
super-agglomerate, rather separated between them and from the core of the
Vel~OB2 association formed by the trio of agglomerates described in
Section~\ref{escorial262829}.   

Escorial~30 might be a spurious agglomeration: 
three of the six stars in the agglomerate have been classified as members
of the concentrated, elongated Pismis~4 cluster (Moffat \& Vogt 1975; Baumgardt
et~al. 2000), while the other three neighbouring early-type stars seem to
populate the outskirt of Pismis~4 and belong to the Vel~OB2 association, or even
to other open clusters in the surrounding area (e.g. \object{HD~72350}~AB in the
van~den~Bergh-Hagen~34 [vdBH~34] cluster; Dias et~al. 2001).
From our data and from Moffat \& Vogt (1975)'s, the Pismis~4 cluster is located
at 500--600\,pc to the Sun, in the background of the Vel~OB2 association.

IC~2391 is, as well as the Pleiades, the Orion Nebula Cluster, $\sigma$~Orionis,
or NGC~2547, a cornerstone for the study of the formation and evolution of stars
and substellar objects at all mass domains (Hogg 1960; Perry \& Hill 1969;
Stauffer et~al. 1989, 1997; Patten \& Simon 1996; Barrado y Navascu\'es,
Stauffer \& Jayawardhana 2004; Boudrealt \& Bailer-Jones 2007; Siegler et~al.
2007). 
Besides, Koester \& Reimers (1985) found a white dwarf in the cluster
\linebreak 
(\object{IC~2391~KR~1} [WD~0839--528]), that could be the youngest one yet
found, together with other white dwarfs identified towards NGC~2451 (that is
described in Section~\ref{escorial2427}). 
See also Platais et~al. (2007) for recent improvements of the basic parameters
of IC~2391 ($d$ = 159.2\,pc, age $\sim$ 40\,Ma, [Fe/H] = +0.06).
The seven stars in the Escorial~31 agglomerate are bright, classical members in
the cluster.
The brightest stars in IC~2391 are $o$~Vel (B3IV, $V$ = 3.60\,mag) and HY~Vel
(B3IV, $V$ = 4.83\,mag).
In Section~\ref{distance.ic2391} we derive a new heliocentric distance to
the~cluster.

Four of the six stars in the Escorial~32 agglomerate were classified as members
of the IC~2395 (vdBH~47, \linebreak 
Collinder~192) cluster.
Of the remaining two stars in agglomerate, one (HD~74273) is well separated (by
almost 1\,deg) from the relatively concentrated cluster and may belong to a
different young population in the area, while the other one (HD~74531~A) has
magnitudes, colours, and spectral type (B2V:) that match the IC~2395
spectro-photometric sequence.
The cluster had been poorly investigated (Lyng{\aa} 1960, 1962; Ruprecht 1966;
van~der~Bergh \& Hagen 1975; J{\o}rgensen \& Westerlund 1988) until the
clarifying work by Clari\'a et~al. (2003).
They derived the angular size \linebreak 
($\sim$19\,arcmin), heliocentric distance ($d$ =
800$\pm$40\,pc), and age (6$\pm$2\,Ma) of IC~2395, and suggested a physical
connection to the Vel~OB1c association (Mel'nik \& Efremov 1995) .
We can not provide additional information or improvement from our data.
The brightest star in IC~2395, that may define the cluster centre, is the
binary ($\rho \approx$ 0.3\,arcsec), variable, early-type (B1.5V + mid-B) star
HX~Vel.

Six of the 10 stars in the Escorial~33 agglomerate were listed as genuine
Trumpler~10 {\em cluster} members by Lyng{\aa} (1960, 1962), while one of the four
remaining stars was also listed as a Trumpler~10 {\em association} member by
de~Zeeuw et~al. (1999). 
According to the latter authors, Trumpler~10 (vbBH~53) is not ``a tight open
cluster, but instead an intermediate age OB association''.
However, the \linebreak 
association/cluster nature of Trumpler~10 is not well understood yet.
On the one hand, Lyng{\aa} tabulated a diametre of only 14\,arcmin for the
cluster; this value contrasts with the $\sim$8\,deg estimated by de~Zeeuw et~al.
(1999) for the association.
The doubt of the ``entity'' being a single cluster had been firstly proposed
by Stock (1984)\footnote{There are additional foreground open clusters
at less than 1\,deg to the highest star concentration of Trumpler~10.
NGC~2671 is an $\sim$80\,Ma-old cluster at $d \sim$ 1700\,pc (Pedreros~2000).
ESO~260--8 is an ultracompact H~{\sc ii} region, apart from a radio (methanol
maser) and {\em IRAS} source, at $d \sim$ 1.3\,kpc (Griffith \& Wright 1993;
Walsh et~al. 1997; Dutra et~al. 2003).}.
Although the original Lyng\aa's clustering of bright stars is evident in
digitized plates (see Fig~\ref{theaggregates.31-33}), the mean heliocentric
distance to the sparse association computed by de~Zeeuw et~al. (1999; $d$ =
366$\pm$23) is consistent with previous determinations of distance to the
cluster (Levato \& Malaroda 1975a; Eggen 1980; Lyng{\aa} \& Wramdemark 1984).
The identification of the Escorial~33 agglomerate also supports the existence of
an OB association of age $\sim$15\,Ma in the area, whose densest part would be
the original Lyng\aa's cluster.
Finally, Kaplan, van Kerkwijk \& Anderson (2007) measured the parallactic
distance to the isolated neutron star RX~J0720.4--3125 ($d$ =
360$^{+170}_{-90}$\,pc), and suggested an origin for it in the Trumpler~10 
association 0.7$^{+0.2}_{-0.2}$\,Ma ago.
Escorial~33/Trumpler~10 deserves further analyses on its stellar content and
basic properties (age, distance, spatial distribution, mass~function).

\subsubsection{Escorial~24 (NGC~2451~AB) and~27 (vdBH~23)} 
\label{escorial2427}

The northernmost agglomerates in the Pup-Vel \linebreak 
super-agglomerate, roughly equidistant from the maximum star densities of
CMa-Pup and Pup-Vel, are Escorial~24 \linebreak 
and~25. 

Escorial~24 is associated to NGC~2451~AB, which is actually the superposition of
two clusters at different heliocentric distances. 
First catalogued by Dreyer (1888), NGC~2451 is one of the ten closest open
clusters to the Sun (see \linebreak 
H\"unsch, Weidner \& Schmitt [2003] for a historical review).
It was firstly studied in detail (as a single cluster) by Feinstein (1966) and
in a series of papers by Williams (1966, 1967b).
Prior to 1990, it was revisited by Levato \& Malaroda (1975b) and other authors
(Andersen \& Reiz 1983; \linebreak 
Eggen 1983; Pastoriza \& Ropke 1983; Maitzen \& Catalano 1986; Gilroy 1989). 
There have been later studies in NGC~2451 on close resolved photometry (Platais
et~al. \linebreak 
2001), photometry and membership (Carrier, Burki \& \linebreak 
Richard 1999), X-ray emission (H\"unsch et~al. 2003), and rotation and lithium
abundance (H\"unsch et~al. 2004).  
The ``binary'' status of NGC~2451~A and B has also been confirmed by many
authors (Baumgardt 1998; R\"oser \& Bastian 1994, and references above).
Robichon et~al. (1999) derived $d$ = 188.7$^{+7.0}_{-6.5}$\,pc for NGC~2451~A,
and determined an age of 35--55\,Ma, consistent within error bars with the more
recent determination by Platais et~al. (2001; 60$\pm$20\,Ma) and H\"unsch et~al.
(2003; 56--80\,Ma). 
NGC~2451~A could be the core of a moving group (the Puppis Moving Group --
R\"oser \& Bastian 1994). 
NGC~2451~B is of a similar age but, in contrast, located further away.
Carrier et~al. (1999) derived $d$ = 358$\pm$22\,pc for NGC~2451~B.
These authors also investigated in detail the exceptional emission-line star
V468~Pup~(B6IVe). 

Our list of 14 stars in Escorial~24 approximately match other lists of bright
blue stars in NGC~2451 (both A and B).
However, we classify the stars in Escorial~24 in three groups:
\begin{itemize}

\item 10 star members in NGC~2451~A, with mean proper motion
($\overline{\mu_\alpha \cos{\delta}}$, $\overline{\mu_\delta}$) =
(--21.3$\pm$1.7, +15.4$\pm$1.0)\,mas\,a$^{-1}$ and heliocentric distance
$\overline{d} \le$ 200\,pc (see Section~\ref{distance.ngc2451a}); 

\item three star members in NGC~2451~B (HD~61899\footnote{We use the Henry
Draper nomenclature instead of d$^{03}$~Pup, since it is easy to misunderstand
with d$^{02}$~Pup (HD~61878~AB), that belongs to NGC~2451~A.}, \linebreak 
HD~62991, and HD~63465), with mean proper motion ($\overline{\mu_\alpha
\cos{\delta}}$, $\overline{\mu_\delta}$) = (--10.2$\pm$0.4,
+7$\pm$3)\,mas\,a$^{-1}$ and heliocentric distance $\overline{d}$
340$\pm$30\,pc; and  

\item star V468~Pup, with $d$ = 430$\pm$40 and ($\mu_\alpha \cos{\delta}$,
$\mu_\delta$) = (--0.84$\pm$0.18, --4.4$\pm$0.2)\,mas\,a$^{-1}$. 
This star, although it was considered as a NGC~2451~B member by Carrier
et~al. (1999), has discordant proper motion and heliocentric distance, and
belongs to a {\em third} background stellar population. 

\end{itemize}

To sum up, the Escorial~24 agglomeration is the superposition of at least two
young open clusters.
The brightest star in the area, c~Pup~AB (a K2.5Ib--II star in NGC~2451~B), has
a colour $B_T-V_T$ = 2.038$\pm$0.017\,mag and was not, therefore, in the input
catalogue of our DBSCAN algorithm.
Based on proper motion and parallax analysis, we confirm non-membership in
NGC~2451~A or~B of the bright ($V$ = 6--8\,mag) stars HD~62559 (F2IV),
HD~63738 (F7V), and HD~63291 (K3II--III). 
Besides, HD~62595 is a G7III star that probably belong to NGC~2451~B.
The four {\em Hipparcos} stars also have colours $B_T-V_T >$ --0.05\,mag.

Likewise, the Escorial~27 agglomerate contains 11 stars.
The two brightest ones are the B1.5-type giant \object{MX~Pup}\footnote{We use the
variable star name MX~Pup instead of r~Pup, since it is easy to misunderstand
with R~Pup (an F9Ia supergiant in NGC~2439).} and the dwarf \object{OS~Pup}.
The agglomerate is to the north of the \object{NGC~2546} cluster and in
southeastern vicinity of the H~{\sc ii} region Gum~10 (Gum 1955; Rodgers,
Campbell \& Whiteoak 1960). 
It was Westerlund (1963) who firstly indicated the existence of an OB
association in the region of the long-period cepheid RS~Pup (F8Ia), that 
is relatively close to Escorial~27.
This association, now called \object{Pup~OB3}, is very wide and contains most of
the stars in our agglomerate.
He estimated an age of only 4\,Ma, consistent with the observed spectral types
and apparent magnitudes.
Later, \linebreak 
van~den~Bergh \& Hagen (1975) reported a not previously catalogued cluster
roughly centred on OS~Pup. 
This cluster, van~den~Bergh-Hagen~23 (vdBH~23), had quite uncertain parameters
until was rediscovered in more recent works (Platais et~al. 1998; Kharchenko
et~al. 2005). 
An heliocentric distance of $d \sim$ 384\,pc was estimated by Dias et~al. (2001).
Although not all the stars in the Escorial~27 agglomerate have been catalogued
as members of vdBH~23, they probably represent the same entity.
The O9.5II supergiant HD~68450 is probably a star in the background and does not
belong to Escorial~27/vdBH~23.
From the weighted mean of the parallaxes of the remaining ten stars, we derive
an heliocentric distance of $\overline{d}$ = 320$\pm$30\,pc, slightly lower than
Dias et~al. (2001)'s estimation.
A dedicated study of Escorial~27/vdBH~23 is still to be carried~out.

\subsubsection{Escorial~25 (P~Puppis)}
\label{escorial25}

Most of the ten stars in the Escorial~25 agglomerate are poorly known. 
Three of them are the $\beta$~Cep variable QS~Pup and the very bright giants
P~Pup~AC and HD~63578 ($V$ = 4.10 and 5.22\,mag, respectively). 
The latter star and other two ones in the agglomerate (HD~63343 and HD~63449~AB
-- a close double with $\rho$ = 0.496\,arcsec) were classified as members of the
Vel~OB2 association by de~Zeeuw et~al. (1999). 
Apart from photometry and astrometry, there is no additional information for the
remaining stars.

The region displays an evident larger surface density of early-type stars than
in its surroundings, and it is located at a considerable angular separation from
the Vel~OB2 trio of agglomerates (Section~\ref{escorial262829}).
Escorial~25 probably belongs to the Pup-Vel super-agglomerate, but not to the
``core'' of Vel~OB2. 
We propose for the first time the existence of an open cluster coinciding with
Escorial~25, approximately centred on P~Pup~AC, the brightest star in the
agglomerate. 
Therefore, we call it the ``P~Puppis cluster''.
The ten {\em Hipparcos} stars in Escorial~25/P~Puppis delineate a \linebreak 
sharp cluster sequence in the $B_T$ vs. $B_T-V_T$  \linebreak 
colour-magnitude diagram.
HD~63007 (B5V) has, however, proper motion and heliocentric distance slightly
different from the remaining cluster member candidates;
HD~63007 might be, therefore, a foreground contaminant.
Accounting for the remaining nine {\em Hipparcos} stars in the cluster, if
confirmed, it would be located at an heliocentric distance of $\overline{d}$ =
470$\pm$70\,pc. 

The only very bright, red star in the area is the evolved giant \object{Q~Pup}
(K0III), with heliocentric distance and proper motion inconsistent with
membership in Escorial~25/P~Puppis ($d$ = 70$\pm$2\,pc, $\mu >$
100\,mas\,a$^{-1}$). 
As a result, there is no evidence of cluster stars in the Red and Asymptotic
Giant Branches.
The brightest star in the cluster is an early-type giant (B0III) that has not
moved away from main-sequence.
The same can be applied to the subgiants QS~Pup and \linebreak 
HD~63578 (B1.5IV).
The next stars in brightness order are B3--5V dwarfs, that have evolutionary
ages of roughly \linebreak 
20\,Ma.
We will assume a very young age of $10^{+10}_{-5}$ for the P~Puppis cluster.

Further analyses, required to ascertain the nature of Escorial~25/P~Puppis, will
be presented in Section~\ref{PPup}.

\section{Discussion}

\subsection{Two new clusters? $\eta$~Orionis and P~Puppis}

\subsubsection{$\eta$~Orionis}
\label{etaOri}

Among the 35 identified agglomerates, two of them have never been proposed to
be real clusters (i.e. their stars are gravitationally bound).
They are the $\eta$~Orionis overdensity and the P~Puppis cluster candidate.

As already mentioned in Section~\ref{escorial4}, the centre of the
$\eta$~Orionis overdensity does not coincide with the massive (double) star
$\eta$~Ori~AB.
In order to avoid the limits of the {\em Hipparcos} catalog and our colour
selection, we carried out a new Virtual Observatory analysis of the region.
In particular, we loaded with Aladin all Tycho-2 stars at less than 2\,deg to
the central coordinates in Section~\ref{escorial4} ($N$ = 734), and selected all
of them with colours $B_T - V_T <$ +0.25\,mag and magnitudes $V_T <$ 10.50\,mag
($N$ = 96).
We checked that all the selected stars have very low proper motions ($\mu \le$
10\,mas\,a$^{-1}$) and follow a relatively narrow secuence in the $B_T$ vs. $B_T
- V_T$ colour-magnitude diagram, consistent with its membership in Orion, and
that redder non-selected stars of similar brightess have colours (and
parallaxes) typical of field dwarfs in the foreground.

We counted the number of selected stars in coronae centred on both $\eta$~Ori~AB
and the survey area central coordinates, and did not find a clear evidence of a
{\em dense} star agglomeration as has been found in other sites in Ori~OB1 (like
in the Orion Nebula Cluster and the $\sigma$~Orionis cluster).
Besides, the elongated shape of the overdensity gets more patent with the
Tycho-2 data, which contrasts with the symmetrical, radial density gradient
found in well-characterized clusters (Cartwright \& Whitworth 2004; Caballero
2008a).
To sum up, we confirmed that a spread overdensity of bright, early-type stars
does exist in the area of $\eta$~Ori~AB in the Ori~OB1a association.
However, we failed to corroborate the existence of a (gravitationally bound)
cluster in that region.

\subsubsection{P~Puppis}
\label{PPup}

\begin{figure*}
\centering
\includegraphics[width=0.48\textwidth]{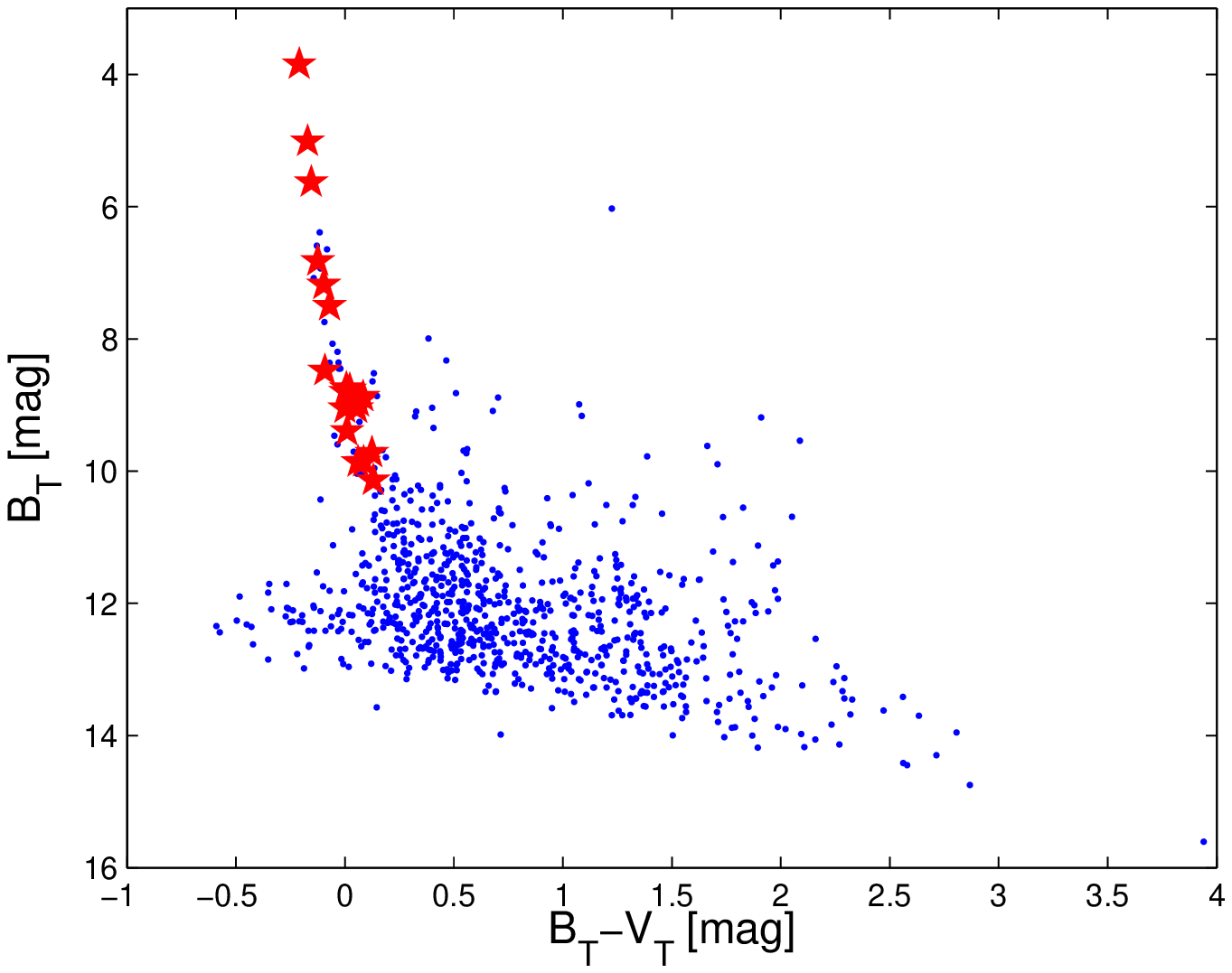} 
\includegraphics[width=0.48\textwidth]{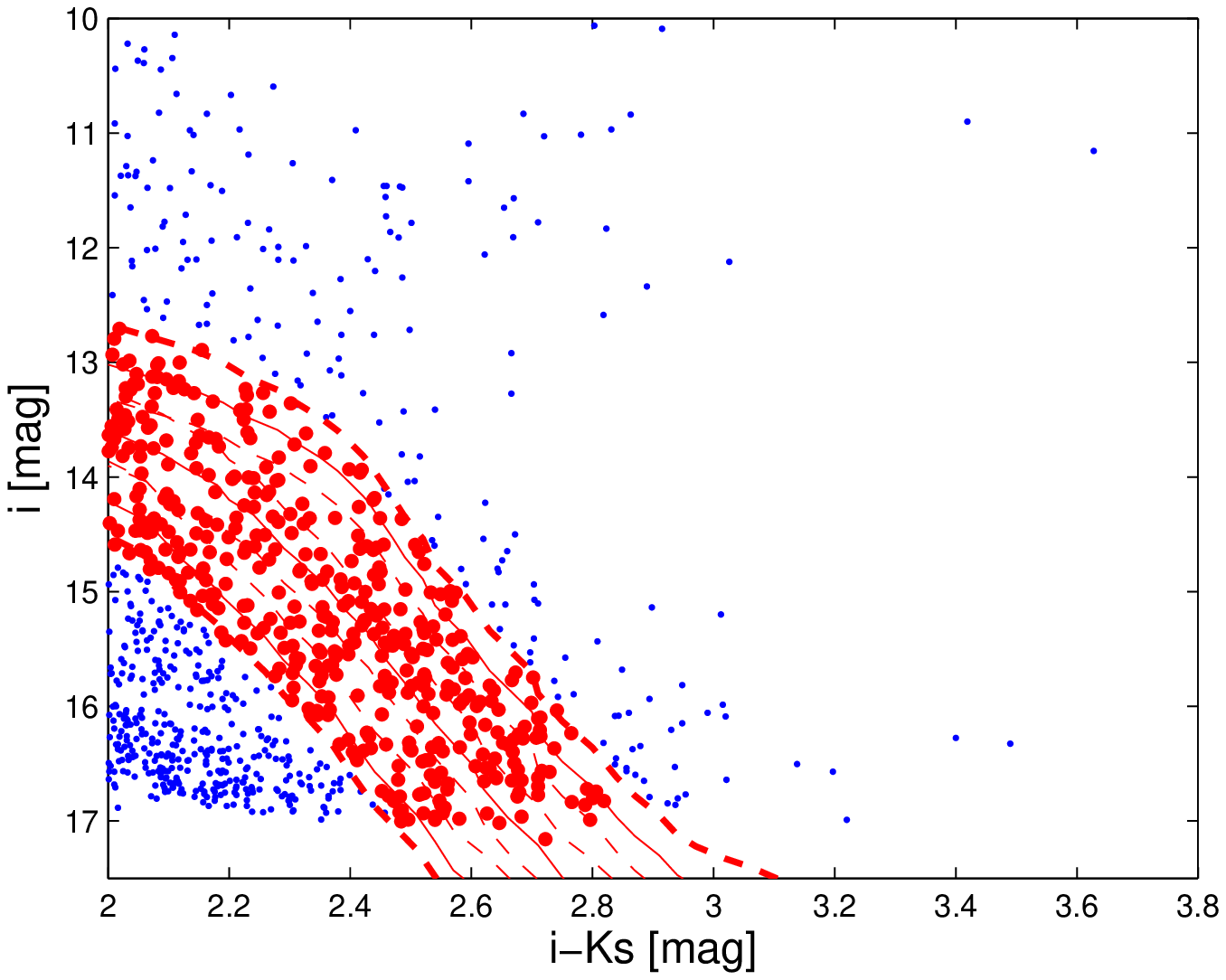} 
\includegraphics[width=0.48\textwidth]{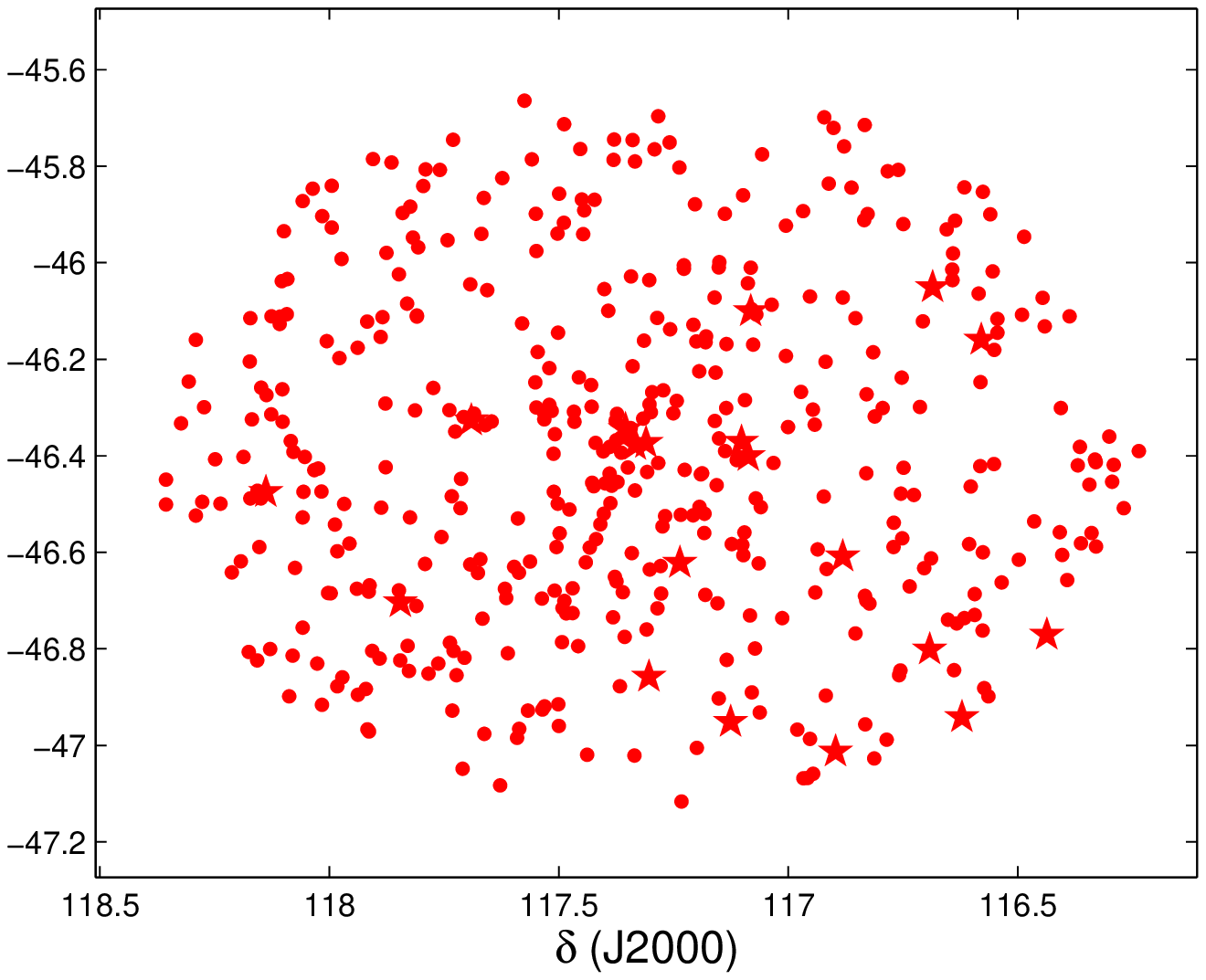} 
\includegraphics[width=0.48\textwidth]{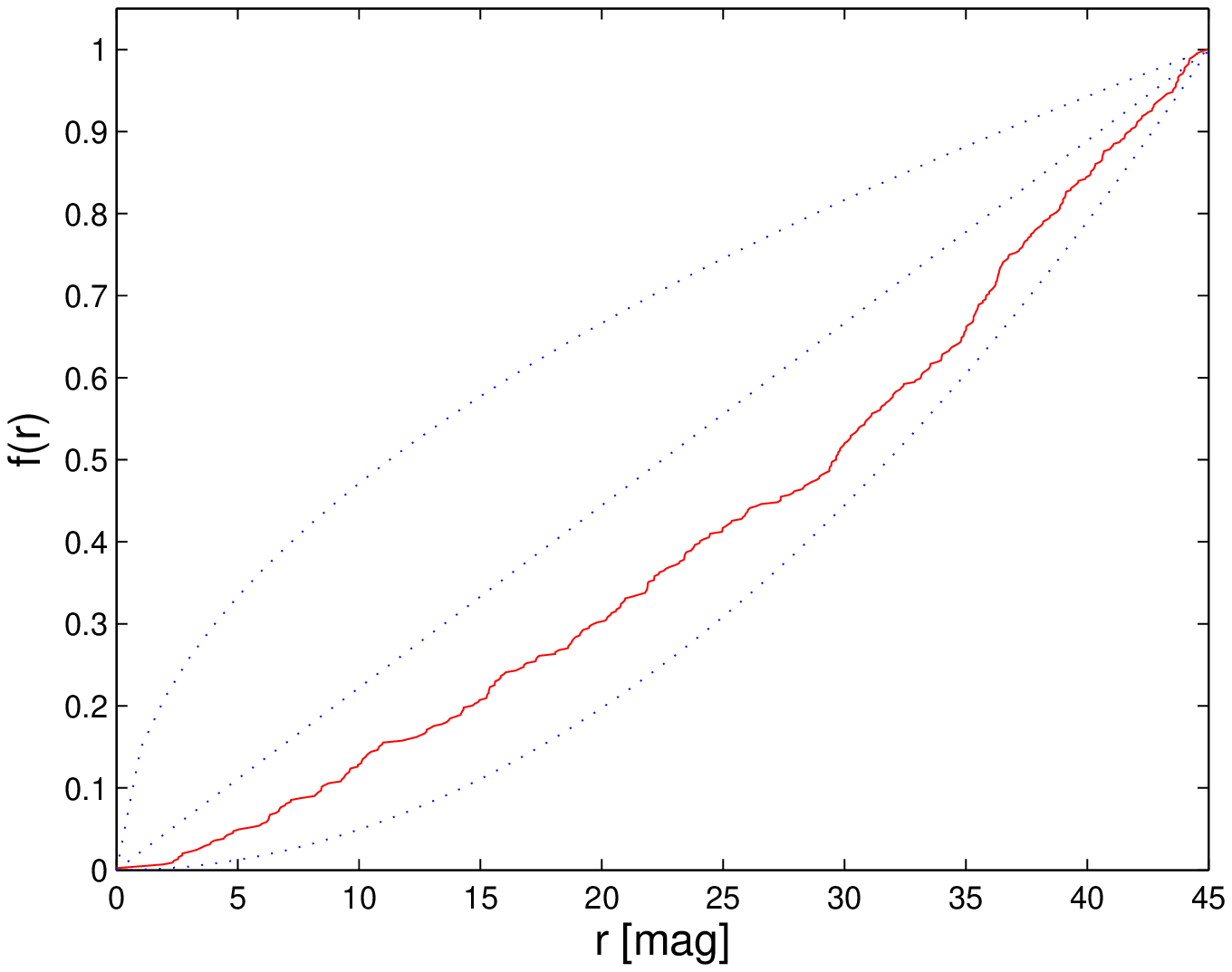} 
\caption{Some diagrams illustrating the follow-up of the P~Puppis cluster.
{\em Top left:} $B_T$ vs. $B_T - V_T$ colour-magnitude diagram showing Tycho-2
cluster member candidates at less than 45\,arcmin to P~Pup~AC (filled --red--
stars) and remaining objects at up to 90\,arcmin to the early-type binary (small
--blue-- dots).
{\em Top right:} the faintest, reddest portion of the $i$ vs. $i-K_{\rm s}$
diagram showing DENIS/2MASS cluster member candidates at less than 45\,arcmin to
P~Pup~AC (filled --red-- circles) and remaining objects at the same angular
separation (small --blue-- dots).
Solid (red) lines are for the 5, 10, and 20\,Ma-old NextGen98 isochrones (from
top to bottom) shifted at $d$ = 470\,pc, while dashed (red) lines are for the
previous isochroness shifted at $d$ = 400 (above) and 570\,pc (below).
{\em Bottom left:} spatial distribution of the Tycho-2 (filled stars) and
DENIS/2MASS (filled circles) cluster member candidates.
{\em Bottom right:} normalized cumulative number, $f(r)$, of cluster member
candidates, in solid (red) line, and theoretical power-law distributions for
$f(r) \propto r^{1/2}$, $r^1$, and $r^2$ (from top to bottom).
Compare with figures in Caballero~(2008a).}   
\label{thePPup}
\end{figure*}

We followed a similar procedure to that of the $\eta$~Orionis overdensity in the
previous section for studying the possible cluster nature of the agglomerate
Escorial~25/P~Puppis. 
First, we selected all Tycho-2 stars satisfying the colour and magnitude
restrictions listed above.
In this case, we only loaded the stars in a 90\,arcmin-radius circle centred on
P~Pup~AC.
The photometric sequence in the \linebreak
colour-magnitude diagram in top left panel of Fig.~\ref{thePPup} is very sharp,
as already noticed in Section~\ref{escorial25} with the {\em Hipparcos} data.
Remarkably, the three brightest stars in the agglomerate (i.e. the two resolved
components in P~Pup~AC and the subgiant QS~Pup) lie on a circle with a radius of
only 45\,arcmin, which subtends and area four times smaller than the
90\,arcmin-radius circle.
Far from expecting four times less blue, bright stars, approximately half of
them are within the smaller radius (i.e. there are twice more blue bright stars
in the inner circle than expected -- the deviation is at the level of five times
the square root of the Poissonian variance), suggesting a {\em dense} star
agglomeration.

We probed the very low-mass stellar population (close to the substellar
boundary) in the 45\,arcmin-radius circle using the Dee Near Infrared Survey of
the Southern Sky (DENIS; Epchtein et~al. 1997) and the Two-Micron All Sky Survey
(2MASS; Skrutskie et~al. 2006) and the NextGen98 theoretical isochrones of
Baraffe et~al. (1998). 
We proceed as in Caballero \& Solano (2008), by loading DENIS and 2MASS data and
cross-matching the sources.
We filtered the faintest ($i >$ 10\,mag), reddest ($i-K_{\rm s}$) sources with
the highest quality photometry ($\delta i$, $\delta J$, $\delta H$, $\delta
K_{\rm s} <$ 0.1\,mag).
Since the Virtual Observatory selection of P~Puppis cluster member candidates
cannot be carried out through the definition of a lower envelope of
spectroscopically confirmed young stars (as in the $\sigma$~Orionis; Caballero
2008c) or of a redder envelope that both maximizes the number of cluster
members and minimizes the number of contaminants (as in the Alnilam/Mintaka
region; Caballero \& Solano 2008), we opted for the classical selection
criterion based on theoretical isochrones.
As can be seen in the colour-magnitude diagram in top left panel in window
Fig.~\ref{thePPup}, the theoretical P~Puppis sequence, considering all possible
uncertainties in the cluster heliocentric distance ($d$ = 470$\pm$70\,pc) and
age ($10^{+10}_{-5}$\,Ma), is quite broad and overlaps with the location of a
lot of (foreground) field M-type dwarfs.
Other colour-magnitude diagrams, performed with the same technique in regions
of exactly the same area and galactic latitude, and of approximately the same
galactic longitude, are quite similar, which indicates that many of the selected
cluster member candidates (between the 5\,Ma, 400\,pc and the 20\,Ma, 540\,pc
NextGen isochrones) are actually late-type interlopers.
Although some elaborated de-contamination technique could be applied (e.g.
Caballero, Burgasser \& Klement 2008), a central overdensity of selected cluster
members is already obvious (see the spatial distribution of selected sources in
the bottom left panel in Fig.~\ref{thePPup}).
Since there are 425 DENIS/2MASS selected sources in the 45\,arcmin-radius
circle, one should expect $\sim$100 such objects in the four-times smaller area
defined by a 22.5\,arcmin-radius circle centred on P~Pup~AC.
In contrast, we found more than 150 of them.
In other words, there are at least $\sim$50 (young) low-mass stars at less than
20--25\,arcmin to the B0 giant that probably form part of the P~Puppis cluster.

Other way of illustrating the central overdensity of \linebreak
P~Puppis is the diagram of the normalized cumulative number of cluster member
candidates, $f(r) = N(r) / N(r_{\rm max})$, as a function of angular distance to
the the centre, $r$. 
A wealth of details on how to compute and interpret it are in Caballero (2008a).
In short, a power-law distribution $f(r) \propto r^2$ corresponds to a uniform
distribution of objects in the survey area, while $f(r) \propto r^1$ corresponds
to a volume density proportional to $r^{-2}$, which is consistent with the
collapse of an isothermal spherical molecular cloud (e.g. Burkert, Bate \&
Bodenheimer 1997).
A distribution with an intermediate index, as that one observed in the Mintaka
cluster (Caballero \& Solano 2008) and in P~Puppis from $r$ = 0 to
$\sim$25\,arcmin (bottom right panel in Fig.~\ref{thePPup}) corresponds to the
superposition of a radially-concentrated cluster with $f(r) \propto r^1$ and a
homogeously-distributed population of (late-type, field dwarf) contaminants.

To sum up, we discover a star cluster surrounding the young, massive P~Pup~AC
giant.
However, a dedicated spectroscopic follow-up will be necessary to disentangle
\linebreak 
between the stellar populations of the cluster and of the field.

\subsection{New sites for substellar searches}
\label{browndwarfs}

Some of the agglomerates identified in our work are valuable ``hunting grounds''
for the search of faint brown dwarfs and planetary-mass objects. 
In particular, the first brown dwarf, \object{Teide~1}, was found in the
Pleiades cluster (Rebolo et~al. 1995), while the Orion Nebula and the
$\sigma$~Orionis clusters harbour the great majority of the planetary-mass
objects with spectroscopy (Lucas \& Roche 2000; Zapatero Osorio et~al. 2000).
Other listed agglomerates that have held substellar searches are the Orion Belt
(surrounding Alnilam and Mintaka), NGC~2264, NGC~2547/$\gamma$~Velorum, and
IC~2391. 
However, there are agglomerates in our list that share properties (youth,
closeness, low extinction) that also make them ideal sites for substellar
searches.

After discarding the possible spurious agglomerates and the groupings associated
to Collinder~121 and Vel~OB2, that are too far away and/or too sparse for
efficient searches, we retained 14 agglomerates for a follow-up inspection of
their parameters.
Of them, five seem to be at prohibitive heliocentric distances larger than
650\,pc (i.e. NGC~2353, Collinder~132, NGC~2362, Pismis~4, and IC~2395).
Besides, M~47, although being located at $d \sim$ 470\,pc, has a Pleiades-like
age and, therefore, their hypothetical brown dwarfs will be about 3\,mag
fainter.

We maintain eight agglomerates as new sites for substellar searches.
Two of them are obvious choices, because they resemble their ``elder brothers''
(in contraposition to ``fraternal twins'') in Orion: 25~Orionis and
$\eta$~Orionis.
Even in the case that $\eta$~Orionis is not a real cluster, the large observed
stellar surface density will facilitate forthcoming studies (stellar and
substellar objects are expected to have proportional surface densities;
Caballero 2008a). \linebreak
NGC~2451~A, with an age of $\sim$45\,Ma and a very close heliocentric distance
of $d$ = 180--200\,pc (see Section~\ref{distance.ic2391}) is other apparent
option.

If compiled basic data for vdBH~23 (Escorial~27) and Trumpler~10 (Escorial~33),
with distances and ages in the approximate intervals 320--360\,pc and 1--4\,Ma,
are correct, then they might be new cornerstones for the substellar \linebreak
searches in the near future.
Finally, there remain in order of interest, NGC~2232 (Escorial~8; $d \sim$
320\,pc, age $\sim$ 30\,Ma), Collinder~140 (Escorial~19; $d \sim$ 370\,pc, age
$\sim$ 20\,Ma), and our new cluster P~Puppis (Escorial~25; $d \sim$ 470\,pc, age
$\sim$ 10\,Ma).
The first two clusters in the trio seem to be IC~2391 twins at larger
distances, while P~Puppis resembles the stellar groupings in the Orion~Belt.

\subsection{Sub-structure of super-agglomerates}
\label{sub-structure}

We find evidence of spatial sub-structure within classical OB associations.
First, the ``isolated cluster'' Escorial~35 could be a physical grouping within
Lac~OB1 \linebreak
(Section~\ref{escorial3435}).
Secondly, our three super-agglomerates \linebreak
(Orion, CMa-Pup, and Pup-Vela) are, by definition, the \linebreak
combination of a number of individual agglomerates. 
There can be sub-sub-structure (i.e. sub-structure within the agglomerates), as
described in, for example, Sections~\ref{escorial9} (NGC~2264), \ref{escorial6}
(Orion Belt), and~\ref{escorial2427} (NGC~2451~AB), but it was not our aim to
detect~it.


The Orion super-agglomerate is composed of five agglomerates, whose boundaries
do {\em not} coincide with classical divisions within the Ori~OB1 complex.
The Orion Sword (Escorial~5) is the superposition of 23 stars previously
ascribed to the sub-associations Ori~OB1c[1--4] and \linebreak 
Ori~OB1d, and includes the brightest stars of, e.g., the Orion Nebula Cluster.
The Orion Belt (Escorial~6) coincides with Ori~OB1b, except for the fact that it
does not contain the stars in the nearby $\sigma$~Orionis cluster, that form the
largest spatial overdensity in a new agglomerate of early-type stars close to
the Horsehead Nebula (Escorial~7).
Finally, we \linebreak 
identify two agglomerates surrounding the very bright stars 25~Ori (Escorial~3)
and $\eta$~Ori (Escorial~4) in the Ori~OB1a sub-association. 
While the 25~Ori agglomerate was identified very recently, we report here for
the first time the existence of a possible open cluster surrounding $\eta$~Ori.

The classical view of sequential formation of subgroups in OB associations by
Elmegreen \& Lada (1977) is reinforced by our results.
As shown in Fig.~\ref{theaggregates.orion}, the stars in the Orion
super-agglomerate are aligned in the north-south direction.
Noticeably, the oldest (or less young) agglomerate is located northward (25~Ori;
$\sim$10\,Ma), the youngest one southward (Orion Sword; $\sim$1\,Ma), and those
ones with intermediate ages (Orion Belt, Horsehead, and, possibly, \linebreak 
$\eta$~Ori) in the middle. 
Extrapolating backward, it might happen that the $\lambda$~Orionis cluster (to
the north of the Orion super-agglomerate) is older than 25~Ori.
On the other hand, the ionization and front shocks produced in the Orion Sword
might also initiate in the near future another cycle of OB star formation to the
south of Orion, where large concentrations of molecular gas have been detected
by the {\em IRAS}~satellite.


We have broken up the CMa-Pup super-agglomerate \linebreak 
into Collinder~132 (Escorial~17) and Collinder~140 (Escorial~19) to the south
and Collinder~121 to the north. 
\linebreak 
Collinder~121 has been splitted, in its turn, into ten agglomerates.
One of them, Escorial~10, would be the original \linebreak 
Collinder~121 compact group that
gives the name to the extended, loose OB association.
The few dedicated spectro-photometric analyses in the region do not allow to
conclude if there are younger and older regions within the \linebreak 
super-agglomerate.
The spatial sub-structure suggests, however, that Collinder~121 is an analog to
the Ori~OB1 association at about 1\,kpc to the~Sun.


The Pup-Vel super-agglomerate displays a much larger spatial distribution
complexity.
Indeed, the large spread on heliocentric distances, from $\overline{d}$ =
147$\pm$6\,pc of IC~2391 to the $d \sim$ 800\,pc of IC~2395 indicates that not
all of them are physically bounded.
NGC~2451~A, at $\overline{d}$ = 181$\pm$9\,pc, would be another open cluster in
the foreground of the extended Vel~OB2 association.
However, the remaining agglomerates in the super-agglomerate could actually have
a common origin.
All of them except Pismis~4 ($d \sim$ 500--600\,pc) lie in a relatively narrow
interval of heliocentric distances:
($i$) vdBH~23 at $\overline{d}$ = 320$\pm$30\,pc;
($ii$) NGC~2451~B at $\overline{d}$ = 340$\pm$30\,pc;
($iii$) Trumpler~10 at $d \sim$ 360\,pc;
($iv$) $\gamma$~Velorum, the core of the Vel~OB2 association, NGC~2547, and
V468~Pup at $d$ = 400--450\,pc; and
($v$) P~Puppis at $\overline{d}$ = 470$\pm$70\,pc.
The mean age of these star clusters is about 30\,Ma, with only vdBH~23 having
been reported to be significatively younger (age $\sim$ 4\,Ma).
Although a careful investigation of the parallactic distances to the stars in
these agglomerates will not be carried out until the {\em GAIA} mission is
operative, the resemblance between the ages and distances of $\gamma$~Velorum
and NGC~2547 (that coincide with the trio of agglomerates Escorial~26, 28,
and~29) suggests that an important fraction of the Vel~OB2 association is
actually the superposition of different star-forming regions {\em \`a~la}~Orion.

\subsection{Early-type stars at $d \le$ 200\,pc}

A bit less than 10\,\% of the investigated OB-type stars are at relatively short
heliocentric distances, at $d \le$ 200\,pc.
Most of them belong to the nearby clusters Pleiades (Escorial~1), NGC~2451~A
(Escorial~24), and IC~2391 (Escorial~31), but there are also possible
contaminants in the foreground of our agglomerates (listed in last column in
Table~\ref{agglomerates}), and errors in the {\em Hipparcos} catalogue.
On the one hand, cluster stars with accurate parallax measurement are important
to derive cluster distances, and compare its spectro-photometric sequence with
theoretical models.
On the other hand, some of the foreground early-type stars may belong to the
rare type of bright B-type dwarfs unbound to known open clusters, which are
generally subject of intensive follow-ups in the literature.
Next, we describe the identified early-types stars at $d \le$~200\,pc.

\subsubsection{Pleiades}
\label{distance.pleiades}

Because of its importance for the ``cosmic distance ladder'', the disagreement
on the actual heliocentric distance to the Pleiades is still an open problem in
Astrophysics (see recent works by Zwahlen et~al. [2004], Soderblom et~al.
[2005], Southworth, Maxted \& Smalley [2005], and Percival, \linebreak 
Salaris \& Groenewegen [2005]).  
Far from taking back again the dispute between the main-sequence fitting and the
{\em Hipparcos} distance moduli, we have computed an additional measurement of
the Pleiades heliocentric distance by \linebreak 
weight-averaging the re-reduced {\em Hipparcos} parallaxes of \linebreak 
the ten stars in our agglomerate Escorial~1. 
The computed distance to the Pleiades, $\overline{d}$ = 120$\pm$5\,pc, is still
shorter than the adopted value of 134--139\,pc derived from isochrone fitting,
but slightly larger than the previous (probably incorrect) {\em Hipparcos}
distance of 118\,pc. 
The two {\em Hipparcos} measurements are, however, consistent with each other
within error bars. 
There is no especial improvement in our $\overline{d}$ if we also add Asterope
and Calaeno, classical Pleiads with {\em Hipparcos} parallax that were not in our
input catalogue (Section~\ref{escorial1}).

\subsubsection{NGC 2451 A}
\label{distance.ngc2451a}

In Section~\ref{escorial2427}, we identified seven {\em Hipparcos} stars as
\linebreak 
members of the NGC~2451~A cluster.
Our weight-averaged heliocentric distance to the cluster, $\overline{d}$ =
181$\pm$9\,pc, is consistent with (although slightly lower than) the value
derived by Robichon et~al. (1999) from the original {\em Hipparcos} data ($d$ =
188.7$^{+7.0}_{-6.5}$\,pc). 
Carrier et~al. (1999) derived, however, $d$ = 197$\pm$12\,pc using Geneve
photometry and isochrone fitting.
Nevertheless, the difference between this value and ours, $\Delta d$ =
16$\pm$15\,pc, could be lower if state-of-the-art theoretical tracks and
improved metallicity and age were used (as in Platais et~al. [2007] for IC~2391
-- see below). 
As a result, there is no hint of NGC~2451~A  displaying a Pleiades-like
disagreement between  heliocentric and isochrone distances.

\subsubsection{IC 2391}
\label{distance.ic2391}

IC~2391 (Section~\ref{escorial30313233}), with an age of about 40\,Ma, is
probably the youngest open cluster in the solar vicinity (see, however, Barrado
y Navascu\'es et~al. [2004] and references therein).
Platais et~al. (2007) have provided the most complete discussion on the problem
of the distance to IC~2391.
They derived an isochrone distance of 159.2\,pc, while the classical {\em
Hipparcos} distance is $d \sim$ 146\,pc.
From the \linebreak 
weight-average of the parallaxes of the seven stars in our Escorial~31
agglomerate, we derive a new {\em Hipparcos} distance of $\overline{d}$ =
147$\pm$6\,pc.
The distance controversy is, therefore, still~standing.

\subsubsection{Early-type field dwarfs}
\label{fielddwarfs200pc}

   \begin{table*}
      \caption[]{Early-type field dwarfs at $d \le$ 200\,pc.} 
         \label{fielddwarfs}
     $$ 
         \begin{tabular}{c l l c c c}
            \hline
            \hline
            \noalign{\smallskip}
HIP		& Name			& Spectral	& $d$		& $M_V$	& $B-V$	\\
 		& 			& type		& [pc]		& [mag] & [mag] \\
            \noalign{\smallskip}
            \hline
            \noalign{\smallskip}
25469		&        {HD 35716}   	& B9		& 180$\pm$30	& 2.21	&--0.03	\\ 
25592		&        {HD 35926} AB  & B7IV+		& 180$\pm$40	& 2.08	&--0.07	\\ 
25600		&        {HD 35957}	& B8		& 190$\pm$40	& 2.07	&--0.05	\\ 
35083		&        {HD 56342}   	& B3V		& 193$\pm$8	&--1.10	&--0.14	\\ 
37173		&        {m Pup} A(C)B  & B8IV++	& 190$\pm$9	&--1.72	&--0.09	\\ 
37229		&      {k$^{01}$~Pup} A & B6V		& 110$\pm$8	&--0.79	&--0.16	\\ 
		&      {k$^{01}$~Pup} B & B5IVn		& 110$\pm$8	&--0.58	&--0.11	\\ %
37304		&        {HD 61687} AB 	& B6V+		& 193$\pm$18	& 0.34	&--0.11	\\ 
39716		&        {HD 67704}  	& A0V		& 200$\pm$14	& 0.45	&--0.06	\\ 
            \noalign{\smallskip}
            \hline
         \end{tabular}
     $$ 
   \end{table*}

In Table~\ref{fielddwarfs}, we list eight foreground agglomerate contaminants at
less than $d \le$ 200\,pc (one binary was actually resolved by Tycho-2 in its
two components).
Apart from shorter heliocentric distances than to the corresponding
agglomerates, they display different proper motions and, in most cases, are
located in abnormal positions in the colour-magnitude diagrams.
Six of them are B6--A0 dwarfs at distances $d \approx$ 180--200\,pc and are not
of especial interest.
\linebreak 
However, the two remaining stars stand~out.

As described in Section~\ref{escorial1719}, {HD~56342} (B3V, $V_r$ =
+33.4$\pm$2.8\,km\,s$^{-1}$, $v \sin{i}$ = 26\,km\,s$^{-1}$) is a contaminant of
the poorly known Collinder~132 cluster.  
The differences between the HD~56342 and cluster heliocentric distances,
\linebreak 
$\Delta d \approx$ 450\,pc, and proper motion, $\Delta \mu \approx$
11\,mas\,a$^{-1}$, make unlikely its membership in Collinder~132.
The star has been spectroscopically investigated in detail by Lyubimkov et~al.
(2002) and subsequent papers.
HD~56342 has typical effective temperature, mass, surface gravity, and helium
abundance for its spectral type and class.
Furthermore, they derived $d$ = 209$\pm$36\,pc based solely on
spectro-photometric parameters, which is consistent with our
parallactic distance ($d$ = 193$\pm$8\,pc). 
Savage et~al. (1985) and Berghoefer, \linebreak 
Schmitt \& Cassinelli (1996) also derived independent distances at  $d \approx$
240 and 236\,pc from ultraviolet interstellar extinction with the Astronomical
Netherlands Satellite and from hydrogen column density with {\em ROSAT}. 
With age and mass of 54$\pm$9\,Ma and 5.3$\pm$0.3\,$M_\odot$, respectively
(Lyubimkov et~al. 2002), HD~56342 is one of the very few early B dwarfs in the
solar neighbourhood that do not belong to a known~cluster.

The case of {k$^{01}$~Pup}~AB (CD--26~4707) is even more extreme.
Its proper motion, of $\mu \sim$ 29\,mas\,a$^{-1}$, is inconsistent with
membership in the Escorial~23 agglomerate, that probably forms part of the
Collinder~121 complex ($d \sim$ 0.5--1.0\,kpc).
{k$^{01}$~Pup}~AB is a binary system of B5--6 stars (HD~61555 and HD~61556:
$\rho$ = 9.913$\pm$0.003\,arcsec, $\theta$ = 318\,deg, $\Delta H_P$ =
0.21$\pm$0.01\,mag; Perryman et~al. 1997) located at only $d$ = 110$\pm$8\,pc.  
This distance agrees with that estimated by Lindroos (1985), at $d \sim$
125\,pc. 
He also derived a very young age, of about 12\,Ma.
The evolved nature of the secondary (class IV), that is a helium variable star
(Rivinius et~al. 2003), may indicate a slightly older age. 
In any case, {\em if} the {\em Hipparcos} parallactic distance is correct
(binarity may have affected the parallax measurement), {k$^{01}$~Pup}~AB could
be the closest massive very young star (age $\le$ 50\,Ma, $M \ge$
5\,$M_\odot$), even closer than the OB-type stars in Upper~Scorpius,
R~Coronae~Australis, $\rho$~Ophiuchi, Chamaeleon~I+II, and IC~2391 ($d$ =
130--150\,pc)\footnote{The most massive nearby star is probably
\object{$\beta$~Pic} (A6V, $d$ = 19.27$\pm$0.19\,pc, age $\sim$ 12\,Ma).
{k$^{01}$~Pup}~AB, roughly contemporary, would be five times further, but would
also be much more massive.}.
We are ignorant of the star-forming region where {k$^{01}$~Pup}~AB was born or
if the binary has an associated young moving~group.

\subsection{Missing agglomerates? Missing stars?}

Obviously, our {\em clustering} algorithm is not able to identify all kinds of
star {\em clusters}, but is mostly sensitive to agglomerates with at least six
bright blue stars separated by less than $R_\epsilon$ (in our case, $R_\epsilon$
= 0.8\,deg).
Missing cluster types are globular (e.g. $\omega$~Cen, M~13) and super star
clusters (e.g. Westerlund~1).
None of their stellar components were in the input catalogue described in
Section~\ref{theinputcatalogue} because of their faintness or reddening in the
optical $B_T V_T$ bandpasses (due, in their turn, to very large heliocentric
distances, evolution towards the red giant and asymptotic giant branches, or
high interstellar extinction).

Some well-known open clusters are also missing.
These lacking open clusters can be divided into two classes:
($i$) clusters with ages in the approximate interval 100--600\,Ma (e.g.
$\alpha$~Persei, NGC~2516, M~35, Prasepe, Hyades), and
($ii$) clusters with ages in the approximate interval 1--10\,Ma (e.g. Serpens,
MBM~12, $\rho$~Ophiuchi, IC~348, $\lambda$~Orionis, $h+\chi$~Persei).
There exist also explanations for their non-detection: 
for example, there is no Hyades member with {\em Hipparcos} colour bluer than
$B_T-V_T$ = 0.0\,mag.
The brightest Hyads, Aldebaran and $\theta^{02}$~Tau, are K5III and A7III
giants, while the bluest Hyad with reliable {\em Hipparcos} photometry,
68~Tau~AB, is an A2IV subgiant.
In other words, because of stellar evolution, the Hyades turn-off point is
redwards of our colour selection criterion in Section~\ref{theinputcatalogue}.
Something similar happens to the other 100--600\,Ma open~clusters.

In contrast, the effect of the post-main sequence evolution is barely detectable
in the 1--10\,Ma-old open clusters. 
However, the early-type stars in the youngest ones (e.g. $\rho$~Ophiuchi,
IC~348) also have red colours for their respective spectral types because of
intra-cluster extinction (there are abundant populations of Class~0 and~I
objects --recently born stars surrounded by thick shells-- in many of these
star-forming regions).  
The remaining young open clusters with relatively low extinction (e.g.
$\lambda$~Orionis) have looser, less abundant, early-type stellar populations
than our 35 agglomerates.
Our clustering algorithm is not sensitive, either, to the detection of sparse
very young associations, like Taurus-Auriga and TW~Hydrae, and low-density
Galactic OB associations, like many of those listed in Garmany \&
Stencel~(1992).

\subsection{Effect of the parametrization on the results}
\label{effectoftheparametrization}

We have investigated how a different choice of parameters of the DBSCAN
algorithm (in particular, $R_\epsilon$) would have affected our results.
On the one hand, Table~\ref{agglomerates0.6} shows the early-type {\em
Hipparcos} stars that belong to agglomerates \linebreak 
when we use $R_\epsilon$ = 0.6\,deg instead of 0.8\,deg.
With the new value, there appear two basic differences: 
($i$) there are only 21 agglomerates, and ($ii$) the agglomerates contain no
more than 11 stars.
For example, Escorial~1 (the Pleiades) now contains only six stars, while many
other agglomerates, including the young open clusters Collinder~132, M~47, and
IC~2395, are not identified at all.
Besides, Escorial~3 (25~Ori) is splitted into two different sub-agglomerates
(Escorial~3a and 3b), which may indicate the presence of \linebreak 
sub-sub-structure in Ori~OB1a.
Finally, HIP~26727 \linebreak 
(Mintaka), that is one of the three bright supergiants in 
Escorial~6 (Orion Belt), is classified now as a member of Escorial~7 (Horsehead
and $\sigma$~Orionis).
To sum up, $R_\epsilon$ = 0.6\,deg gives an incomplete sampling of our
aggregates. 

On the other hand, further increasing $R_{\epsilon}$ to 1.0\,deg results in the
fusion of some aggregates, the detection of some new clusters, and the growth of
nearby agglomerates \linebreak 
(Fig.~\ref{thegalaxy}).
In~particular:

\begin{itemize} 
\item 6 agglomerates remain unchanged with respect to the $R_\epsilon$ =
0.8\,deg parametrization (Escorial~1 [Pleiades], 2 [spurious], 4
[$\eta$~Ori], 9 [NGC~2264], 27 [vdBH~23], and~32 [IC~2395]).  

\item 14 agglomerates ``capture'' nearby bright blue stars and increase in size.
Ten of the agglomerates accrete only three or less stars.
There are, however, three agglomerates that drastically grow up, capturing 8
(Escorial~3 --25~Ori--), 9 (Escorial~16 --NGC~2232--), and up to 12 bright blue
stars (Escorial~26 --$\gamma$~Velorum and \linebreak 
NGC~2547--).

\item 15 agglomerates are fused into five larger entities (with additional
capture of nearby bright blue stars).
The resultant groupings are the combination of Escorial~6 \linebreak 
and~7 (Orion Belt and Horsehead in the Orion super-agglomerate), 10--15~and~17,
18~and~20, 22~and~23 (all agglomerates associated to Collinder~121 along with
\linebreak 
Collinder~132), and 28~and~29 (in the dense trio in \linebreak 
Vel~OB2; Section~\ref{escorial262829}).  
As a result, using $R_{\epsilon}$ = 1.0\,deg, the CMa-Pup super-agglomerate is
splitted into only \linebreak four groupings: Collinder~140 (Escorial~19, the
only agglomerate in CMa-Pup that maintain its independence) and the three
fusions in Collinder~121. 
The new Orion division gives the same number of groupings (25~Ori, $\eta$~Ori,
Orion Sword, and ``Orion Belt+Horsehead'').
Remarkably, the large amassment Escorial~10--15~and~17 contains 135 bright blue
stars, a value that does not differ very much from the number of such stars in
\linebreak
de~Zeeuw et~al. (1999)'s Collinder~121 asociation.

\item 19 agglomerates are new.
All of them, except three, contain six or seven stars and have surface densities
that are $\pi 0.8^2 / \pi 1.0^2$ = 0.64 times smaller than our spurious
agglomerate Escorial~2.
Many of them are probably spurious agglomerates as well, and they would require
a careful follow-up, as that carried out for the original parametrization.
Two of the remaining agglomerates lie on CMa-Pup-Vel region, with N$_\star$ = 8
and 9, and may belong to the sparse stellar population of the local spiral arm
of the Galaxy (Section~\ref{results}).
Finally, there is a new isolated agglomerate with a relative large number of
stellar components, N$_\star$ = 10, and separated from the large overdensities
in Fig.~\ref{thegalaxy}.
It includes the B0Vp-type star \object{$\theta$~Car}, that is the brightest one
of the IC~2602 open cluster (age $\sim$ 30\,Ma, $d$ = 135$\pm$9\,pc -- Randich
et~al. 1995; Stauffer et~al.~1997). 

\end{itemize} 

   \begin{table*}
      \caption[]{Agglomerates of early-type {\em Hipparcos} stars
       with the alternative parameters $N_{\rm MinPts}$ = 6 and $R_\epsilon$ =
       0.6\,deg.} 
         \label{agglomerates0.6}
     $$ 
         \begin{tabular}{l l l}
            \hline
            \hline
            \noalign{\smallskip}
No.	& Original 	& {\em Hipparcos} stars			\\
	& agglomerate 	& (HIP)					\\
	& (Escorial) 	& 					\\
            \noalign{\smallskip}
            \hline
            \noalign{\smallskip}
I	& 1    		& 17499, 17527, 17531, 17573, 17608, 17702 \\
II	& 2    		& 23279, 23287, 23295, 23328, 23473, 23508 \\
III	& 3a   		& 25163, 25235, 25288, 25302, 25340, 25469 \\
IV	& 3b   		& 25241, 25378, 25411, 25533, 25567, 25582, \\
	&      		& 25592, 25648, 25655, 25751, 25752 \\
V	& 4    		& 25293, 25394, 25480, 25552, 25557, 25600 \\
VI	& 5    		& 26197, 26199, 26241, 26314, 26345, 26427 \\
VII	& 6    		& 26106, 26213, 26311, 26319, 26334, 26405, \\
	&      		& 26439, 26464, 26508, 26683 \\
VIII	& 7    		& 26549, 26551, 26579, 26656, 26694, 26713, \\
	&      		& [26727] \\
IX	& 8    		& 30580, 30660, 30700, 30758, 30761, 30772, \\
	&      		& 30789 \\
X	& 9    		& 31917, 31951, 31955, 31978, 32030, 32053 \\
XI	& 10   		& 33062, 33070, 33087, 33165, 33208, 33215, \\
	&      		& 33276, 33309, 33410 \\
XII	& 11   		& 33586, 33666, 33695, 33721, 33796, 33841 \\
XIII	& 12   		& 33935, 33970, 34041, 34048, 34167, 34227 \\
...	& 13   		& ... \\
...	& 14   		& ... \\
...	& 15   		& ... \\
...	& 16   		& ... \\
...	& 17   		& ... \\
XIV	& 18   		& 35267, 35370, 35412, 35461, 35503, 35539, \\
	&      		& 35597 \\
XV	& 19   		& 35700, 35761, 35795, 35822, 35855, 35905, \\
	&      		& 36045 \\
...	& 20   		& ... \\
...	& 21   		& ... \\
...	& 22   		& ... \\
...	& 23   		& ... \\
...	& 24   		& ... \\
XVI	& 25   		& 37926, 37953, 38020, 38028, 38159, 38164 \\
XVII	& 26   		& 39873, 39919, 40011, 40016, 40024, 40059 \\
XVIII	& 27   		& 40218, 40255, 40268, 40274, 40321, 40324 \\
...	& 28   		& ... \\
XIX	& 29   		& 40662, 40742, 40825, 40851, 40872, 40921 \\
...	& 30   		& ... \\
XX	& 31   		& 42400, 42459, 42504, 42535, 42536, 42715, \\
	&      		& 42726 \\
...	& 32   		& ... \\
XXI	& 33   		& 43055, 43085, 43182, 43209, 42340, 43326, \\
	&      		& 43392 \\
...	& 34   		& ... \\
...	& 35   		& ... \\
            \noalign{\smallskip}
            \hline
         \end{tabular}
     $$ 
   \end{table*}

\subsection{Possible errors in the input catalogue}
\label{possiblehipparcoserrors}

Apart from the large number of {\em Hipparcos} stars without parallax
determination (indicated with ellipses ``...'' in the $d$ column in
Table~\ref{theaggregates}), we have also identified two stars that suffered from
systematic errors during the {\em Hipparcos} reduction.
Both of them display proper motions with very large error bars and improbable
distances of a few tens AU (indicated with square brackets ``[~]'' in the $d$
column in Table~\ref{theaggregates}).
They are the binary HIP~23279 + HIP~23287 (\object{HD~32039} +
\object{HD~32040}, B9Vn, in Escorial~2; $d$ = 34$\pm$9\,pc) and HIP~35503
(\object{HD~57281}~AB, B5V, in Escorial~18; $d$ = 70$\pm$30\,pc).
We have not taken them into account in the analysis.

Finally, it seems that there was a misunderstanding in the Bonner Durchmusterung
and/or the {\em Hipparcos} \linebreak 
catalogues between two 10--11th-magnitude stars separated by $\sim$57\,arcsec. 
The actual K2-type star BD+05~1825 \linebreak 
(TYC~189-1314-1), presented in Section~\ref{theinputcatalogue}, is the
easternmost and brightest one of the pair at passbands \linebreak 
$V_T R I J H K_{\rm s}$.
The westernmost star (HIP~38575, TYC-189-1700-1), that is brighter than
BD+05~1825 only at passband $B_T$, is the only {\em Hipparcos} star in the area.
HIP~38575 has typical colours of late B or early A~dwarfs.

\section{Summary}

We have used the DBSCAN (Density-Based Spatial Clustering of Applications with
Noise) data clustering algorithm to identify spatial agglomerates (``clusters'')
of {\em Hipparcos} stars with colours $B_T-V_T <$ --0.05\,mag.

A total of 35 agglomerates of early-type stars (with \linebreak 
spectral types late O, B, and very early A) have arosen from the search.
They are ascribed to young open clusters and OB associations, except for a few
of them whose physical grouping is uncertain and seem to be spurious detections
(Escorial~2 and, possibly, Escorial~16, 30, 34, and 35).
Of the remaining agglomerates, four are associated to known open clusters and
dense star-forming regions (the Pleiades [Escorial~1], NGC~2264 [Escorial~9],
M~47 [Escorial~21], and the poorly known NGC~2232 cluster [Escorial~8]), while
26 form three super-agglomerates (agglomerates of~agglomerates). 

The Orion super-agglomerate, that coincides with a \linebreak 
large fraction of the classical Orion~OB1 complex, is splitted into five
agglomerates: 25~Orionis (Escorial~3), $\eta$~Orionis (Escorial~4), Orion Sword
(Escorial~5), Orion Belt (Escorial~6), and Horsehead (Escorial~7).
This division is atypical: the most important differences with classical
divisions are the existence of a new {\em overdensity} of stars surrounding
\linebreak 
$\eta$~Ori~AB, and the membership in a population different from the Orion Belt
of the  $\sigma$~Orionis cluster.
We also confirm the recently identified cluster around 25~Ori.

The CMa-Pup super-agglomerate is broken up into \linebreak 
Collinder~132 (Escorial~17), Collinder~140 (Escorial~19),
and Collinder~121, which is separated for the first time into ten agglomerates
(Escorial~10--15, 18, 20, 22, and~23).
The Pup-Vel super-agglomerate is a conglomeration of open \linebreak 
clusters in the foreground (at $d <$ 200\,pc: IC~2391 [Escorial~31], NGC~2451~A
[Escorial~24]), background (at $d >$ 500\,pc: IC~2395
[Escorial~32] and, possibly, Pismis~4 [Escorial~30]), and at the average
distance to the Lac~OB2 association (at $d$ = 340--470\,pc).
The latter association may comprise different young stellar populations
associated to, e.g., Trumpler~10, $\gamma$~Velorum, and NGC~2547.
Many of the agglomerates discussed here need, however, careful
spectro-photometric analyses.

The open cluster P~Puppis, presented here for the first time, could also be a
distant member of the Lac~OB2 association.
We have carried out a dedicated study of the cluster using Tycho-2, DENIS, and
2MASS data and theoretical isochrones of the Lyon group and quantified its
radial density gradient using the normalized cumulative number of cluster member
candidates. 

We have listed seven agglomerates whose substellar \linebreak
populations will probably be investigated in the future.
They are the 25~Orionis and $\eta$~Orionis overdensities in the \linebreak 
Ori~OB1a association, the nearby, $\sim$45\,Ma-old cluster \linebreak 
NGC~2451~A, and four poorly known clusters at $d$ = 320--470\,pc with younger
ages (NGC~2232, P~Puppis, vdBH~23, and Trumpler~10). 

We have investigated in detail the early-type {\em Hipparcos} stars in
agglomerates with heliocentric distances $d \le$ 200\,pc.
By weight-averaging their parallaxes, we have computed new distances to the
Pleiades ($\overline{d}$ = 120$\pm$5\,pc),\linebreak NGC~2451~A ($\overline{d}$
= 181$\pm$9\,pc), and IC~2391 ($\overline{d}$ = 147$\pm$6\,pc). 
The dispute between the isochrone fitting and {\em Hipparcos} distances
still goes on for the Pleaides and IC~2391, but there is no controversial for
NGC~2451~A. 
The remaining ten stars at $d \le$ 200\,pc are two stars with errors in the
{\em Hipparcos} parallax measurements, six B6--A0 dwarfs with typical absolute
magnitudes and colours in the foreground, and two standing early/intermediate B
stars. 
The latter stars are HD~56342 (age $\sim$ 54\,Ma, $M \sim$ 5.3\,$M_\odot$), a
B3V star at $d$ = 193$\pm$8\,pc, and {k$^{01}$~Pup}~AB, two B5--6 dwarfs
separated by $\rho \sim$ 1090\,AU and located at only  $d$ = 110$\pm$8\,pc to
the Sun.
If the {\em Hipparcos} parallax measurement is correct, then {k$^{01}$~Pup}~AB
would be the closest massive very young star (age $\le$ 50\,Ma, $M \ge$
5\,$M_\odot$).

Finally, we have also discussed which is the effect of the parameter choice
($N_{\rm MinPts}$ and $R_\epsilon$) on our agglomerate identification and why
some known clusters (e.g. Hyades) have not been~identified.

Many of the results and hypotheses presented here will be corroborated, refined,
or corrected by the future ESA {\em GAIA} space mission and subsequent analyses,
including \linebreak more sophisticated data clustering~algorithms.

\acknowledgements
Partial financial support was provided by the Universidad Complutense de Madrid,
the Spanish Virtual Observatory and the Spanish Ministerio Educaci\'on y Ciencia 
under \linebreak grants AyA2005--02750, AyA2005--04286 and AyA2005--24102--E of
the Programa Nacional de Astronom\'{\i}a y Astrof\'{\i}sica and by the Comunidad
Aut\'onoma de Madrid under PRICIT project S--0505/ESP--0237 (AstroCAM). 
LD also ackowledges financial support from Ministerio Educaci\'on y Ciencia
through grants Consolider MOSAICO and FIS04--271.
This research has made use of the SIMBAD, operated at Centre de Donn\'ees
astronomiques de Strasbourg, France, and the NASA's Astrophysics Data System as
bibliographic service.
We acknowledge the use of NASA's {\em SkyView} facility 
({\tt http://skyview.gsfc.nasa.gov}) \linebreak located at NASA Goddard Space
Flight~Center. 

\appendix

\section{Early-type {\em Hipparcos} stars in agglomerates}
\label{imagesandbigtable}

{\em NOTE to the reader: Figs. \ref{theaggregates.1-3}
to~\ref{theaggregates.34-35} and the complete Table~\ref{theaggregates} will
appear in the on-line version of Astronomische Nachrichten.}  

\begin{figure*}
\centering
\caption{False-colour DSS-2 images of the agglomerates of early-type {\em
Hipparcos} stars centred on reference stars in Table~\ref{agglomerates}.
Blue, green, and red are for photographic $B_J$, $R_F$, and $I_N$.
North is up, east is left.
All the images are 3$\times$3\,deg$^2$ in size.
Fits images were obtained from {\em SkyView} and combined with {\em DS9}.
From left to right: Escorial~1 (Pleiades), 2, and~3 (25~Ori).}
\label{theaggregates.1-3}
\end{figure*}
%

\begin{figure*}
\centering
\caption{Same as Fig.~\ref{theaggregates.1-3}, but for Escorial~4 ($\eta$~Ori;
{\em left}), Escorial~5 (Orion Sword; {\em centre}) and
Escorial~6 (Orion Belt; {\em right}).}  
\label{theaggregates.4-6}
\end{figure*}
%

\begin{figure*}
\centering
\caption{Same as Fig.~\ref{theaggregates.1-3}, but for the agglomerates
Escorial~7 (Horsehead; {\em left}), Escorial~8 (NGC~2232; {\em centre}), and
Escorial~9 (NGC~2264; {\em right}).}
\label{theaggregates.7-9}
\end{figure*}
%

\begin{figure*}
\centering
\caption{Same as Fig.~\ref{theaggregates.1-3}, but for the agglomerates
Escorial~10, 11, and~12, from left to right.
They are associated to Collinder~121.}
\label{theaggregates.10-12}
\end{figure*}
%

\begin{figure*}
\centering
\caption{Same as Fig.~\ref{theaggregates.1-3}, but for the agglomerates
Escorial~13, 14, and~15, from left to right.
They are also associated to Collinder~121.}
\label{theaggregates.13-15}
\end{figure*}
%

\begin{figure*}
\centering
\caption{Same as Fig.~\ref{theaggregates.1-3}, but for the agglomerates
Escorial~16 (including NGC~2353; {\em left}), Escorial~17 (Collinder~132; {\em
centre}), and Escorial~18 (associated to Collinder~121; {\em right}).}
\label{theaggregates.16-18}
\end{figure*}
%

\begin{figure*}
\centering
\caption{Same as Fig.~\ref{theaggregates.1-3}, but for the agglomerates
Escorial~19 (Collinder~140; {\em left}), Escorial~20 (associated to
Collinder~121; {\em centre}), and Escorial~21 (M~47; {\em right}).}
\label{theaggregates.19-21}
\end{figure*}
%

\begin{figure*}
\centering
\caption{Same as Fig.~\ref{theaggregates.1-3}, but for the agglomerates
Escorial~22 and~23 (associated to Collinder~121) and Escorial~24 (NGC~2451~AB;
{\em right}).} 
\label{theaggregates.all.22-24}
\end{figure*}
%

\begin{figure*}
\centering
\includegraphics[width=0.33\textwidth]{a25.eps} 
\caption{Same as Fig.~\ref{theaggregates.1-3}, but for the agglomerates
Escorial~25 (P~Puppis; {\em left} -- {\em NOTE to the reader: this is the
only agglomerate to be plotted}), Escorial~26 (associated to
Vel~OB2, and including NGC~2547 and $\gamma$~Velorum; {\em centre}), and
Escorial~27 (vdBH~23; {\em right}).} 
\label{theaggregates.25-27}
\end{figure*}
%

\begin{figure*}
\centering
\caption{Same as Fig.~\ref{theaggregates.1-3}, but for the agglomerates
Escorial~28 and~29 (associated to Vel~OB2), and Escorial~30 (including vdBH~34
and Pismis~4; {\em right}).}
\label{theaggregates.28-30}
\end{figure*}
%

\begin{figure*}
\centering
\caption{Same as Fig.~\ref{theaggregates.1-3}, but for the agglomerates
Escorial~31 (IC~2391; {\em left}), Escorial~32 (IC~2395; {\em centre}), and
Escorial~33 (Trumpler~10; {\em right}).} 
\label{theaggregates.31-33}
\end{figure*}
%

\begin{figure*}
\centering
\caption{Same as Fig.~\ref{theaggregates.1-3}, but for the agglomerates
Escorial~34 and~35.}
\label{theaggregates.34-35}
\end{figure*}

\begin{table*}
\caption{\label{theaggregates} 
Early-type {\em Hipparcos} stars in aggregates {\em (PARTLY shown)}.}  
\begin{tabular}{l l l l c c c c c}
\hline
\hline
\noalign{\smallskip}
Agglomerate 		& HIP		& Name			& Spectral	& $d$		& $\alpha$	& $\delta$	& $\mu_\alpha \cos{\delta}$	& $\mu_\delta$		\\
	 		& 		& 			& type		& [pc]		& (J2000)	& (J2000)	& [mas\,a$^{-1}$]		& [mas\,a$^{-1}$]	\\
\noalign{\smallskip}
\hline
\noalign{\smallskip}
\object{Escorial 1} 	& 17499		&        {Electra}	& B6IIIe	& 124$\pm$4	& 03 44 52.54	& +24 06 48.0	& +20.9$\pm$0.3			& --46.0$\pm$0.2	\\
 			& 17527		&        {18 Tau}	& B8V		& 125$\pm$6	& 03 45 09.74	& +24 50 21.3	& +20.4$\pm$0.4			& --46.5$\pm$0.4	\\
 			& 17531		&        {Taygeta}	& B6IV		& 125$\pm$5	& 03 45 12.49	& +24 28 02.2	& +21.3$\pm$0.4			& --40.5$\pm$0.4	\\
 			& 17573		&        {Maia}		& B8III		& 117$\pm$4	& 03 45 49.61	& +24 22 03.9	& +21.2$\pm$0.4			& --46.0$\pm$0.3	\\
 			& 17608		&        {Merope}	& B6IVe		& 116$\pm$5	& 03 46 19.57	& +23 56 54.1	& +21.2$\pm$0.4			& --43.6$\pm$0.3	\\
 			& 17702		&        {Alcyone}	& B7IIIe	& 124$\pm$6	& 03 47 29.08	& +24 06 18.5	& +19.3$\pm$0.4			& --43.7$\pm$0.3	\\
 			& 17776		&        {HD 23753}	& B8V		& 118$\pm$5	& 03 48 20.82	& +23 25 16.5	& +18.0$\pm$0.4			& --46.6$\pm$0.3	\\
 			& 17847		&        {Atlas} AB	& B8III+	& 117$\pm$5	& 03 49 09.74	& +24 03 12.3	& +17.7$\pm$0.4			& --44.2$\pm$0.3	\\
 			& 17851		&        {Pleione}	& B8IVev	& 117$\pm$4	& 03 49 11.22	& +24 08 12.2	& +18.1$\pm$0.3			& --47.2$\pm$0.3	\\
 			& 17900		&        {HD 23923}	& B8V		& 115$\pm$8	& 03 49 43.53	& +23 42 42.7	& +16.7$\pm$0.6			& --44.8$\pm$0.5	\\
\object{Escorial 2} 	& 23279		&        {HD 32039}	& B9Vn		&     ...	& 05 00 32.54	& +03 36 53.3	&(+10.5$\pm$1.3)		&(--18.2$\pm$1.3)	\\ 
		 	& 23287		&        {HD 32040}	& B9Vn		& [34$\pm$9]	& 05 00 33.93	& +03 36 56.7	&(+13.0$\pm$1.4)		&(--16.1$\pm$1.3)	\\ 
		 	& 23295		&        {HD 32056}	& B9		& 220$\pm$70	& 05 00 39.79	& +03 15 55.2	&  +0.4$\pm$1.3			&  --4.6$\pm$0.8	\\ 
		 	& 23328		&        {V1360 Ori}	& B8V		& 260$\pm$40	& 05 01 06.02	& +03 43 02.4	& --2.0$\pm$0.7			&  --6.7$\pm$0.4	\\ %
		 	& 23473		&        {HD 32359}	& B9V		& 500$\pm$200   & 05 02 44.55	& +03 27 27.7	&  +2.5$\pm$0.8			&  --3.8$\pm$0.6	\\ %
			& 23508		&        {HD 32431}	& B8V		&     ...	& 05 03 17.10	& +04 00 56.0	&  +0.5$\pm$1.0			&   +3.8$\pm$0.7	\\ %
\object{Escorial 3} 	& 25163		&        {HD 35177}	& B9		& 350$\pm$150	& 05 23 01.93	& +01 41 48.9	& --1.0$\pm$1.2			&  --1.1$\pm$0.6	\\ 
		 	& 25164		&        {HD 35194}	& B9		& 270$\pm$70	& 05 23 01.96	& +00 26 20.6	& --1.3$\pm$0.9			&  --3.0$\pm$0.6	\\ 
		 	& 25179		&        {HD 35203}	& B6V		& 320$\pm$90	& 05 23 10.15	& +01 08 22.6	& --2.7$\pm$1.0			&  --3.2$\pm$0.5	\\ 
		 	& 25235		&        {V1156 Ori}	& B3Vw		&     ...	& 05 23 50.36	& +02 04 55.8	&  +0.4$\pm$1.0			&   +0.9$\pm$0.5	\\ 
		 	& 25241		&        {HD 35305}	& B6.5IV/V	& 380$\pm$130   & 05 23 51.38	& +00 51 46.3	&  +3.1$\pm$0.9			&  --0.8$\pm$0.5	\\ 
		 	& 25288		&        {HD 35407} 	& B4IVn		& 470$\pm$160   & 05 24 36.10	& +02 21 11.4	& --0.2$\pm$0.7			&  --0.9$\pm$0.4	\\ 
		 	& 25302		&        {25 Ori}	& B1Vpe		& 320$\pm$90	& 05 24 44.83	& +01 50 47.2	&  +0.2$\pm$0.7			&  --0.2$\pm$0.4	\\ 
		 	& 25340		&        {HD 35501} AB 	& B8V+		&     ...	& 05 25 11.40	& +01 55 24.0	&  +1.0$\pm$1.1			&  --2.9$\pm$0.5	\\ 
		 	& 25378		&        {HD 35588} AB  & B2.5V+	& 380$\pm$60	& 05 25 47.02	& +00 31 12.9	&  +1.2$\pm$0.4			&  --0.8$\pm$0.2	\\ 
		 	& 25411		&        {HD 35612}   	& B8		& 320$\pm$100	& 05 26 06.00	& +00 50 02.4	&  +2.0$\pm$0.9			&  --1.8$\pm$0.5	\\ 
		 	& 25469		&        {HD 35716}   	& B9		& 180$\pm$30	& 05 26 48.11	& +02 04 05.9	& --2.3$\pm$1.1			&   +0.1$\pm$0.5	\\ 
		 	& 25533		&        {HD 35834} AB  & B8V+		& 370$\pm$150	& 05 27 36.88	& +01 06 27.3	& --0.2$\pm$1.1			&   +2.1$\pm$0.5	\\ 
		 	& 25567		&        {HD 35881}   	& B8V		& 210$\pm$50	& 05 27 54.23	& +01 06 18.2	&  +1.0$\pm$1.0			&  --0.2$\pm$0.5	\\  
		 	& 25582		&        {HD 35912}   	& B2V		& 400$\pm$90	& 05 28 01.47	& +01 17 53.7	& --0.9$\pm$0.6			&   +0.7$\pm$0.3	\\ 
		 	& 25592		&        {HD 35926} AB  & B7IV+		& 180$\pm$40	& 05 28 10.12	& +00 47 14.0	& --1.7$\pm$1.3			&  --3.2$\pm$0.6	\\ 
		 	& 25648		&        {HD 36013}   	& B3V:n		& 320$\pm$60	& 05 28 45.29	& +01 38 38.2	& --1.0$\pm$0.6			&  --0.2$\pm$0.3	\\ 
		 	& 25655		&        {V1372 Ori}   	& B5Vne		& 390$\pm$90	& 05 28 48.46	& +02 09 53.0	& --1.7$\pm$0.5			&  --0.6$\pm$0.3	\\ 
		 	& 25751		&        {HD 36166}   	& B2V		& 440$\pm$70	& 05 29 54.77	& +01 47 21.3	&  +1.3$\pm$0.4			& --1.33$\pm$0.18	\\ 
		 	& 25752		&        {HD 36165}   	& B7V		& 230$\pm$50	& 05 29 55.55	& +02 08 31.8	& --2.0$\pm$0.8			&   +0.4$\pm$0.4	\\ 
		 	& 25881		&        {HD 36392}   	& B3V		& 250$\pm$60	& 05 31 29.89	& +01 41 24.1	& --1.9$\pm$1.2			&   +0.1$\pm$0.5	\\ 
			& 25897		&        {HD 36429}   	& B5V		& 200$\pm$30	& 05 31 41.22	& +02 49 58.2	&  +2.6$\pm$0.8			&  --1.3$\pm$0.4	\\ 
			& 25979		&        {HD 36549}	& B6Vwp		& 320$\pm$130	& 05 32 39.49	& +02 05 31.8	&  +3.5$\pm$1.3			&  --0.9$\pm$0.5	\\ 
			& 26098		&        {HD 36741}	& B2V		& 340$\pm$60	& 05 33 57.59	& +01 24 27.5	&  +2.8$\pm$0.6			&  --0.6$\pm$0.2	\\ 
...			& ...		& ...			& ...		& ...		& ...		& ...		& ...				& ...			\\ 
			& ...		& ...			& ...		& ...		& ...		& ...		& ...				& ...			\\ 
\noalign{\smallskip}
\hline
\end{tabular}
\end{table*}

\end{document}